\documentclass[12pt]{article}
\usepackage{cite,epsfig,amssymb,amsmath, graphicx,color,subfigure}
\usepackage{multirow}
\evensidemargin \oddsidemargin

\DeclareMathOperator{\sech}{sech}
\usepackage{varioref}

\begin{document}

\vspace{5mm}
\begin{center}
{{\Large \bf Supersymmetric Inhomogeneous Field Theories\\ in 1+1 Dimensions}}
\\[17mm]
O-Kab Kwon$^1$, ~~Chanju Kim$^2$, ~~Yoonbai Kim$^1$\\[2mm]

{\it $^1$Department of Physics,~BK21 Physics Research Division,\\
	Autonomous Institute of Natural Science,\\and Institute of Basic 
	Science, Sungkyunkwan University, Suwon 16419, Korea}\\
{\it $^2$Department of Physics, Ewha Womans University, Seoul 03760, Korea} \\[2mm]
{\it  okab@skku.edu, ~cjkim@ewha.ac.kr, ~yoonbai@skku.edu}
\end{center}
\vspace{15mm}

\begin{abstract}
We study supersymmetric inhomogeneous field theories in 1+1 dimensions which
have explicit coordinate dependence. Although translation symmetry is broken,
part of supersymmetries can be maintained. In this paper, we consider 
the simplest inhomogeneous theories with one real scalar field, which possess
an unbroken supersymmetry. The energy is bounded from below by the 
topological charge which is not necessarily nonnegative definite. 
The bound is saturated if the first-order Bogomolny equation is satisfied. 
Non-constant static supersymmetric solutions above the vacuum involve in 
general a zero mode although the system lacks translation invariance. 
We consider two inhomogeneous theories obtained by deforming supersymmetric
sine-Gordon theory and $\phi^6$ theory. They are
deformed either by overall inhomogeneous rescaling of the superpotential
or by inhomogeneous deformation of the vacuum expectation value. 
We construct explicitly the most general supersymmetric solutions and obtain 
the BPS energy spectrum for arbitrary position-dependent deformations.
Nature of the solutions and their energies depend only on the boundary
values of the inhomogeneous functions. The vacuum of minimum energy is not 
necessarily a constant configuration. In some cases, we find a one-parameter
family of degenerate solutions which include a non-vacuum constant solution
as a special case.
\end{abstract}

\newpage
 
\section{Introduction}

Supersymmetry is usually studied as an extension of Poincare symmetry, 
but it can exist in theories where Poincare symmetry is not relevant
or is partially broken. The latter appears if the theory has 
explicit coordinate dependence in the action.
Such supersymmetric theories have been actively investigated recently.
Janus Yang-Mills theories in four dimensions, where the gauge coupling 
depends on space, are dual to dilatonic deformations of AdS$_5$
space~\cite{Bak:2003jk} and have been generalized to supersymmetric
theories~\cite{Clark:2005te, DHoker:2007zhm, DHoker:2006qeo, Kim:2008dj,
Kim:2009wv, Gaiotto:2008sd, Gaiotto:2008ak}. In three and four dimensions,
mass-deformed ABJM and super Yang-Mills theories admit further deformations 
with inhomogeneous mass functions while some of supersymmetries are 
maintained~\cite{Kim:2018qle, Kim:2019kns, Arav:2020obl, Kim:2020jrs}. 
Gravity duals of those models were also discussed in \cite{Gauntlett:2018vhk, 
Arav:2018njv, Ahn:2019pqy, Hyun:2019juj, Arav:2020asu, Arav:2020obl}. 
In two and three dimensions, supersymmetric theories with impurities
were studied extensively~\cite{Hook:2013yda, Tong:2013iqa, Adam:2019yst,
Adam:2018pvd, Adam:2018tnv, Adam:2019djg, Manton:2019xiq, Adam:2019xuc,
Adam:2019hef}. Supersymmetric boundary/defect integrable models 
in two dimensions are also a subject of active research~\cite{Inami:1995np,
Nepomechie:2001qr,Gomes:2006en, Bowcock:2003dr, Gomes:2007pj}.

Supersymmetric field theories with Poincare invariance show many nice
properties. For example, the energy is nonnegative 
definite~\cite{Iliopoulos:1974zv, Witten:1978mh, Witten:1981nf}. 
In particular, the vacuum energy is 
zero if and only if it is annihilated by each of the supercharges. 
This directly comes from the fact that the Hamiltonian
is the sum of the absolute squares of the supercharges. 

Another important aspect of supersymmetric theories is
the existence of BPS objects~\cite{Prasad:1975kr, Bogomolny:1975de}
which preserve part of the supersymmetries of the theory. 
Imposing appropriate conditions on the supersymmetric parameter, 
one can obtain the first-order Bogomolny equation~\cite{Bogomolny:1975de}
which also minimizes the energy for a given topological sector. 
Classically, the solutions are lumps such as kinks, vortices, or monopoles,
and the moduli space of the solutions reflects the underlying symmetry
of the system.

In this paper, we will address these issues in supersymmetric theories
where the translation symmetry is explicitly broken. For simplicity, 
we will restrict our interest only to two-dimensional classical supersymmetric 
theories with one real scalar multiplet. 
If the theories are Poincare-invariant, 
they have two supercharges in two dimensions.
In case that the superpotential depends explicitly on space, one of the 
supersymmetries is explicitly broken but the other can still be maintained 
with the help of an additional term in the Lagrangian.
We analyze the superalgebra of such case and show that the energy is bounded 
from below by the topological charge which can be negative. 
The bound is saturated if the Bogomolny equation is satisfied. 
Since it is a first-order differential equation, the general solution 
should have an integration constant. On the other hand, the energy
of the solutions turns out to depend only on the boundary values of the 
superpotential. Therefore, the energy is independent of the integration 
constant and hence we can have a one-parameter family of supersymmetric 
solutions above the vacuum, i.e., the solution possesses a zero mode even 
though the system has no translation invariance.
For a kink solution in usual homogeneous theories, the integration constant 
represents the position of the kink. For inhomogeneous theories 
considered in this paper, however, we will see that the physical 
interpretation of the constant depends on the nature of the solutions.

There would be, in principle, infinitely many different ways of deformation. 
In this paper, we consider two types of spatial dependence to the 
superpotential: an overall inhomogeneous rescaling of the superpotential
and inhomogeneous deformation of the vacuum expectation. 
To be specific, we consider supersymmetric sine-Gordon (SSG) theory for the
former deformation, and $\phi^6$ theory for the latter. 
Sine-Gordon theory is one of the most well-studied theories in $1+1$ 
dimensions and possesses many nice properties. It is particularly 
well-suited for our purpose in that it has an overall mass 
parameter in the potential which would be made inhomogeneous. Moreover, it 
has infinitely degenerate vacua and nontrivial 
solutions such as kinks and breathers. We will see how these features are
affected by the inhomogeneous rescaling.
We would like to mention that inhomogeneous rescaling of the potential has 
been studied in the context of BPS-impurity model especially for 
$\phi^4$ theory in~\cite{Adam:2019djg}. We summarize and discuss the results
of $\phi^4$ theory in our context in the appendix.

For the latter inhomogeneity, we investigate $\phi^6$ theory with a 
$\mathbb{Z}_2$ symmetry $\phi \rightarrow -\phi$.
It can have three degenerate minima, one at $\phi=0$ and the
others at nonzero values.
Thus it is the simplest theory having both broken and unbroken vacua with
$\mathbb{Z}_2$ symmetry. By varying the nonzero vacuum expectation values
while keeping the minimum at $\phi=0$ intact, we can study how the inhomogeneity
affect the spectrum of supersymmetric solutions.  We will see that,
with this kind of inhomogeneity, the minima of the potential or even the total 
number of minima depend on the position, and the vacuum profile needs not 
be constant in general, even though $\phi=0$ remains a legitimate 
supersymmetric solution with vanishing energy. This is to be contrasted with
the former deformation where the vacuum is always given by constant 
configurations since the overall shape of the potential remains unchanged 
asymptotically.

For the deformed SSG and $\phi^6$ theories,
even though they contain arbitrary position-dependent 
functions, we are able to construct explicitly the most general 
solutions of the Bogomolny equations.
The BPS energy spectrum does not depend on the details of the inhomogeneity, 
but we will investigate how it changes according to the boundary values 
of the superpotential at spatial infinities. For the theories considered
here, there are either two or three energy levels depending on the 
boundary values of the inhomogeneous function and
the vacuum energy is negative. When there are two energy levels, 
the energy of non-constant solutions becomes degenerate with a constant
solution and they together form a continuously deformable one-parameter 
family of solutions.

The rest of the paper is organized as follows. In section~\ref{sec2}, 
we construct the supersymmetric Lagrangian with one real scalar
multiplet in two dimensions where the superpotential depends on space.
Then, from the superalgebra, we obtain the energy bound. We show
that there is no universal lower bound in energy and that there exists
a zero mode for non-constant supersymmetric solutions above the vacuum. 
In section~\ref{sec3}, we consider theories obtained by inhomogeneous rescaling 
of the superpotential for the SSG theory. We explicitly solve 
the Bogomolny equation for an arbitrary inhomogeneous function and obtain 
the BPS energy spectrum. In section~\ref{sec4}, we investigate
inhomogeneous deformation of the vacuum expectation value in $\phi^6$ theory.
We conclude in section~\ref{sec5} with a brief discussion on 
future research directions. In the appendix, we briefly discuss the
inhomogeneous rescaling of $\phi^4$ theory.

\section{Supersymmetric Inhomogeneous Field Theories}
\label{sec2}

We begin with a brief review of two-dimensional $\mathcal{N}=1$ supersymmetric
theories with a real scalar field. 
The action reads~\cite{ DiVecchia:1977nxl, Hruby:1977nc}, 
\begin{align}\label{2dsQFT}
S =\int d^2x \Big[ -\frac{1}{2} \partial_{\mu} \phi \partial^{\mu} \phi + i \bar{\psi} \gamma^\mu \partial_{\mu} \psi + i W''(\phi) \, \bar{\psi} \psi - \frac{1}{2} W'(\phi)^2 \Big],
\end{align}
where $\phi$ is a real scalar field, $\psi$  a real Majorana fermion,
$W(\phi)$ the superpotential, and $W' \equiv \frac{d W}{d\phi}$.
Here we follow the convention of the spinor indices in the Appendix of 
Ref.~\cite{Kim:2018qle}.   
 
The theory \eqref{2dsQFT} is invariant under the supersymmetric transformation,
\begin{align}\label{susyvar}
&\delta\phi =  i  \psi \epsilon, 
\nonumber\\
& \delta \psi = -\frac{1}{2} \gamma^\mu \partial_{\mu} \phi \,\epsilon + \frac{1}{2} W' \epsilon,
\end{align}
where the supersymmetric parameter $\epsilon$ is a two-component real spinor
and the gamma matrices  are given by $\gamma^{\mu} = (i \sigma^2, \sigma^1)$ with $\mu = 0,1$.
The supercurrent is 
\begin{align}
J_\epsilon^\mu = -i \partial_{\nu} \phi\,\bar{\psi} \gamma^\mu \gamma^\nu \epsilon + i W' \bar{\psi}\gamma^\mu \epsilon,
\end{align}
which leads to two supercharges,
\begin{align}\label{tsc}
 Q_\epsilon = \int dx J_\epsilon^0 =  i\epsilon_+ Q_+ +i  \epsilon_- Q_- \quad {\rm with} \quad   Q_{\pm} &= \int dx \Big( (\partial_{0}\phi \pm \partial_{1}\phi)\psi_{\pm} \mp W' \psi_{\mp} \Big),
\end{align}
where we set $\bar\epsilon^\alpha = (\epsilon_+, \epsilon_-)$ with $\bar \epsilon \equiv \epsilon^\dagger = \epsilon^T$.
They satisfy the superalgebra,
\begin{align}\label{Qpm}
\{Q_{\pm}, Q^\dagger_{\pm} \} =2( P^0 \mp P^1),\qquad \{Q_{\pm}, Q_{\mp}^\dagger \}  =2 T,
\end{align}
where $P^\mu$ denotes the momentum and $T$ is the central charge of the
superalgebra which is identified as the topological charge~\cite{Witten:1978mh}
\begin{align}\label{topc}
T =  \int dx( \partial_{1}\phi) W'(\phi) =\int dx \frac{d W(\phi(x))}{dx} =   W(\phi (\infty)) - W(\phi(-\infty)).
\end{align}
Accordingly, the energy can be written as
\begin{align}\label{homE}
E = P^0 = \frac{1}{4}\{Q_+ \pm Q_-, Q_+^\dagger  \pm Q_-^\dagger\} \mp T.
\end{align}
It implies~\cite{Witten:1978mh} 
\begin{align}\label{EgeT}
E \ge |T|,
\end{align}
which is manifestly nonnegative definite. 
In particular, the vacuum energy cannot be negative. 

We would like to make a few remarks which seem trivial but will be relevant
later. First, the nonnegativity of the energy in
supersymmetric theories comes from the fact that it is bounded from below 
by both $T$ and $-T$ in two different ways. Each bound is saturated if
the corresponding supercharge vanishes. Second, $T$ vanishes 
identically for constant configurations. 
This term is usually neglected when considering the vacuum energy
since the vacuum is given by a constant configuration.
We will later see how these features may be evaded in inhomogeneous
theories which still have unbroken supersymmetries.

Now we consider the case that the superpotential depends explicitly
on the position $W(\phi, x)$. For example, it may occur if some
parameters in the superpotential depend on the coordinate $x$.
Suppose that the supersymmetry transformation rule \eqref{susyvar} remains 
unchanged except that the derivative of the superpotential $W'$ is replaced by
the partial derivative $ \partial W/ \partial \phi$ accordingly,
\begin{align} \label{susyvar2}
\delta \psi = -\frac12 \gamma^\mu \partial_{\mu} \phi \,\epsilon
+ \frac12 \frac{\partial W}{\partial \phi}\epsilon.
\end{align}
The action \eqref{2dsQFT} is no longer invariant under this
transformation but half of the supersymmetry can be preserved by adding
a term in the Lagrangian,
\begin{align}\label{Im2dQFT}
\mathcal{L} = -\frac12 \partial_\mu \phi \partial^\mu \phi
    + i \bar\psi \gamma^\mu \partial_\mu \psi
    + i \frac{\partial^2 W}{\partial \phi^2} \bar\psi \psi
    - \frac12 \left(\frac{\partial W}{\partial \phi}\right)^2
    \mp \frac{\partial W}{\partial x}.
\end{align}
It is straightforward to see that this theory is indeed invariant under
the transformation \eqref{susyvar2} if $\epsilon$ satisfies
\begin{equation}\label{proj}
\gamma^1 \epsilon = \pm\epsilon,
\end{equation}
which implies $\epsilon_- = \pm \epsilon_+$. 
Similar projections for the supersymmetric parameters have been introduced in
supersymmetric quantum field theories with inhomogeneous deformations in 
various dimensions~\cite{DHoker:2006qeo,Kim:2008dj, Kim:2018qle, Adam:2019yst,
Kim:2020jrs}. In particular, Ref.~\cite{Adam:2019yst}
considered two-dimensional theories in the context of BPS preserving
impurity models which we will compare with our approach later in this section.

For clarity, we choose the upper sign in the projection \eqref{proj} from 
now on. Then the corresponding unbroken supercharge is given by  
\begin{align}
 \bar Q_\epsilon = i \epsilon_+ \bar Q
\end{align}  
with 
\begin{align}
\bar Q = \int dx \Big[ (\partial_0\phi 
        + \partial_1\phi - \partial_\phi W)\psi_+
        - (\partial_0 \phi - \partial_1 \phi + \partial_\phi W) \psi_- \Big]. 
\end{align}
By adding the last term in the Lagrangian in \eqref{Im2dQFT}, we now have a 
position-dependent scalar potential
\begin{equation} \label{vphix}
V(\phi,x) \equiv \frac12 \left(\frac{\partial W}{\partial \phi}\right)^2
    + \frac{\partial W}{\partial x},
\end{equation}
which, in particular, need not be nonnegative definite.
Then one can easily obtain
\begin{align}\label{Erel}
E = \frac{1}{4} \{ \bar Q,\, \bar Q^\dagger \} + T,
\end{align}
where the topological charge $T$ is defined in \eqref{topc} with
\begin{align}
\frac{dW}{dx} = \frac{\partial W}{\partial x}
              + \frac{\partial W}{\partial \phi}\frac{d \phi}{dx}.
\end{align}
This can also be directly seen by rewriting the bosonic part of the energy as
\begin{align}
E &= \int dx \left[ \frac12 (\partial_0 \phi)^2 + \frac12 (\partial_1\phi)^2
    + \frac12 \left(\frac{\partial W}{\partial\phi}\right)^2
    +\frac{\partial W}{\partial x} \right] 
    \nonumber \\
  &= \int dx \left[ \frac12 (\partial_0 \phi)^2 
    + \frac12 \left( \partial_1 \phi
             - \frac{\partial W}{\partial\phi}\right)^2 \right] + T.
\end{align}
The vacuum configuration is then time-independent, $\partial_0 \phi=0$,
and satisfies the first-order Bogomolny equation~\cite{Bogomolny:1975de}
\begin{equation} \label{veq}
\frac{d\phi}{dx} - \frac{\partial  W}{\partial \phi} = 0,
\end{equation}
as well as $\psi = 0$. Equation \eqref{veq} coincides with the Killing spinor 
equation $ \delta \psi = 0 $, as it should be. It is also easy to check 
that every static solution of the Bogomolny equation satisfies automatically
the second-order Euler-Lagrange equation. Moreover, the pressure density 
$T_{11}$, which is the spatial diagonal 11-component of the energy-momentum 
tensor, is given by
\begin{align} \label{t11}
T_{11} &= \frac12 (\partial_0 \phi)^2 + \frac12 (\partial_1\phi)^2
    - \frac12 \left(\frac{\partial W}{\partial\phi}\right)^2
    -\frac{\partial W}{\partial x}  \nonumber \\
 &= -\frac{\partial W}{\partial x},
\end{align}
where the Bogomolny equation \eqref{veq} for static BPS configuration
is used in the second line. Thus the pressure density 
of the static solution of \eqref{veq} does not vanish but is given by the 
extra potential term, which is expected since the momentum is not conserved 
in inhomogeneous theories. In fact, it is easy to show that the momentum 
conservation is modified to
\begin{equation}
\partial_0 T^{01} + \partial_1 T^{11} = - \frac{\partial^2 W}{\partial x^2}.
\end{equation}
For static configurations, the first term in the left-hand side should vanish
and hence this equation is consistent with \eqref{t11}.

In the conventional theory in which $W$ does not depend explicitly on
the spatial coordinate $x$, 
the Bogomolny equation \eqref{veq} is trivially satisfied by a constant
configuration $\phi = \phi_0$ with $W'(\phi_0) = 0$. 
Then the central charge vanishes, $T=0$, and the vacuum energy also
vanishes. In inhomogeneous cases, however, $W(\phi,x)$ has nontrivial
dependence on $x$ and there is no reason that the vacuum 
is given by a constant field configuration. 
Note also that the energy is bounded from below only by the topological
charge $T$ itself as in \eqref{Erel} but not by its absolute value $|T|$.   
Therefore there is no guarantee that the energy is nonnegative definite 
even when the supersymmetry is unbroken. 

There can be other solutions of \eqref{veq} with higher energies
above the vacuum. Then the supercharge of these solutions will vanish
and the energy is again equal to the topological charge. Note, however, that
\eqref{veq} is a first-order differential equation and hence the solution
will in general have an integration constant, say $c$, unless the
solution is a constant configuration.
Since the boundary values that $\phi$ can take at spatial infinities are 
in general discrete, the topological charge of the solutions are also discrete. 
Then the energy cannot depend on $c$ except some special case that 
$c$ alters the boundary values in some limit. 
In other words, variation of the integration constant $c$ should be a zero 
mode of the solution. Therefore the existence of the zero mode is a generic 
phenomenon for non-constant supersymmetric solutions above the vacuum even 
if the translation invariance of the theory is explicitly broken\footnote{The
existence of a zero mode has also been discussed in \cite{Adam:2019yst}
in supersymmetric BPS soliton-impurity models.}.
In the following we will see how these are realized in explicit examples.

So far we have assumed that the kinetic term has the standard form as in 
\eqref{Im2dQFT}. 
If we allow a more complicated form in the kinetic term including the
coordinate dependence, we can generalize \eqref{Im2dQFT} to
\begin{equation} \label{Im2dQFT2}
        \mathcal{L} = -\frac12 K^2 \partial_\mu \phi \partial^\mu \phi
        + i \bar\psi \gamma^\mu \partial_\mu \psi
        + \frac{i}{K} \frac\partial{\partial\phi}\left(
           \frac1K \frac{\partial W}{\partial\phi} \right) \bar\psi \psi
        - \frac12 \left(\frac1K \frac{\partial W}{\partial \phi}\right)^2
        \mp \frac{\partial W}{\partial x},
\end{equation}
where $K=K(\phi, x)$ is a function depending on both $\phi$ and $x$ and
cannot be absorbed by a field redefinition. It is straightforward to check 
that this is invariant under the supersymmetric transformation,
\begin{align}\label{susyvar3}
\delta\phi &=  i  \psi \epsilon, \nonumber\\
\delta \psi &= -\frac12 K^2 \gamma^\mu \partial_{\mu} \phi \,\epsilon
        + \frac12 \frac{\partial W}{\partial\phi} \epsilon,
\end{align}
with \eqref{proj}. The energy is again bounded from below by $T$ and the
bound is saturated if 
\begin{equation}
        K \frac{d \phi}{dx} \mp \frac1K \frac{\partial W}{\partial\phi} = 0.
\end{equation}
In this paper, we will only consider the case $K=1$ below.

A few remarks are in order. Supersymmetry in inhomogeneous theories in two
dimensions was also considered in \cite{Adam:2019yst} to incorporate 
the BPS preserving impurity \cite{Adam:2018pvd,Adam:2018tnv} 
into supersymmetric theories. The most general bosonic Lagrangian considered in 
\cite{Adam:2019yst} is
\begin{equation} \label{adamL2}
        \mathcal{L} = -\frac12 H^2 \partial_\mu \phi \partial^\mu \phi
        - \frac{U}{H^2} - G^2 - 2 \frac{\sqrt{U}G}H
        - \sqrt2 H G \frac{\partial \phi}{\partial x},
\end{equation}
where $H$ and $G$ are functions of $\phi$ and $\sigma$ such that 
$H \rightarrow 1$ and $G \rightarrow 0$ when $\sigma \rightarrow 0$.
In our formulation, identifying $K$ in \eqref{Im2dQFT2} with $H$ and
$W = \sqrt2 \int(\sqrt{U} + HG)d\phi$, we can readily see that
the bosonic part of \eqref{Im2dQFT2} reduces to \eqref{adamL2} up to
a total derivative $\frac{d}{dx}\int HG\,d\phi$. The separation of $U(\phi)$
and $G(\phi,\sigma)$ in \cite{Adam:2019yst} is physically motivated 
to represent $G$ as the impurity effect. From the viewpoint of general
inhomogeneous field theories, however, such separation is not needed;
the position-dependent superpotential $W(\phi,x)$ (and $K(\phi,x)$) 
is enough to incorporate impurities as well as more general 
inhomogeneous cases which we will consider in this paper.

Closing this section, we would like to note that the value of
the energy in inhomogeneous case has no absolute meaning just as in usual 
nonsupersymmetric theories. This is because one can always add a 
field-independent term $W_1(x)$ to the superpotential which has
no physical effect,
\begin{equation}
W(\phi,x) \longrightarrow W(\phi,x) + W_1(x).
\end{equation}
Then the overall energy of the theory would be shifted by 
$\Delta E =W_1(\infty) - W_1(-\infty)$ and hence the energy
can be made to take any value with a suitable choice of $W_1$. 
In particular, there is no universal lower bound of energy.
In the following we will not add any field-independent term to the 
superpotential, i.e., we assume that every term in $W$ depends nontrivially 
on $\phi$, which would then provide a natural choice of energy in many 
cases\footnote{This however does not completely fix the ambiguity. See 
section~\ref{sec3}.}.
Also, one can add a total derivative term $\frac{d}{dx}F(\phi(x),x)$ 
to the Lagrangian as seen above, which would shift the energy 
by $F(\phi(\infty),\infty)-F(\phi(-\infty),-\infty)$. Then in theories
with nontrivial boundary values of the field, the energy spectrum can change
by adding this term, since solutions belonging to different topological
sectors would give different energy shifts. This kind of ambiguity, however,
exists even in usual homogeneous theories. We fix the ambiguity
by requiring that there would be no linear term in $\partial_x\phi$ as
in usual homogeneous theories, and also that the energy should coincide with
that of the usual expression in the homogeneous limit. 
In any case, it would not produce any real physical effect ultimately.
In particular, in the present case, it is evident that the central 
charge $T$ is also shifted by the same amount
in \eqref{Erel}, cancelling the energy shift.

\section{Inhomogeneous Rescaling of Superpotential in Supersymmetric
Sine-Gordon Theory}
\label{sec3}

Having constructed the general inhomogeneous supersymmetric Lagrangian,
we now discuss some specific class of inhomogeneities and their consequences.
In this section we consider the theory obtained by inhomogeneous
rescaling of the superpotential. Thus it can be applied to any superpotential. 
The same inhomogeneity has been considered by 
\cite{Adam:2019djg,Manton:2019xiq} in the
context of BPS-impurity model and detailed analysis has been performed with
$\phi^4$ theory in particular for the dynamical aspects~\cite{Adam:2019xuc}.
In this section, we would like to discuss the energy spectrum of supersymmetric 
solutions and consider the deformation of the SSG theory.

Given a superpotential $W_0(\phi)$, let 
\begin{equation} \label{w0}
        W(\phi,x)=g(x) W_0(\phi)
\end{equation}
be the deformed superpotential where $g(x)$ is a position-dependent
rescaling parameter. We assume that it goes to some nonzero finite 
values at spatial infinities so that the interaction does not vanish
at infinities. Then the corresponding deformed potential is
\begin{equation} \label{vtilde}
V(\phi,x)
=\frac12 g^2(x) \left(\frac{dW_0}{d\phi}\right)^2 + g'(x) W_0(\phi).
\end{equation}
The Bogomolny equation \eqref{veq} becomes
\begin{equation} \label{veq2}
        \phi'(x) - g(x) W'_0(\phi) = 0.
\end{equation}
Under the coordinate transformation from $x$ to $G(x) = \int g(x) dx$, 
\eqref{veq2} reduces to the undeformed form. Thus the rescaling of the
superpotential is always solvable irrespective of an arbitrary
function $g(x)$ \cite{Adam:2019djg}.
Note that the boundary condition for $\phi(x)$ at spatial infinities 
is the same as that of the undeformed theory, i.e.\
they are extremum points of $W_0(\phi)$. 

Let $\phi_i$ be extremum points of $W_0(\phi)$, and denote the boundary values
of $g(x)$ as $g(\infty)=g_R$ and $g(-\infty)=g_L$. Then the constant
configurations $\phi = \phi_i$ are obvious solutions of \eqref{veq2}
and the corresponding energies are 
\begin{align} \label{egi}
E_{\phi=\phi_i} &= T  \nonumber \\
      &= g(\infty) W_0(\phi(\infty))
                  - g(-\infty) W_0(\phi(-\infty)) \nonumber \\
      &= (g_R - g_L) W_0(\phi_i).
\end{align}
The vacuum energy is the minimum of these energies, 
$E_\textrm{vac} = \min_{\phi_i} E_{\phi=\phi_i}$.
Note that, in this expression, the value of superpotential at the vacuum 
$W(\phi_i)$ appears explicitly in the vacuum energy due to inhomogeneity
of the parameter $g(x)$.

As an example, let us now consider the SSG theory~\footnote{Sine-Gordon 
theory with a somewhat different inhomogeneity has been considered in 
\cite{Adam:2019hef} in the context of BPS-impurity model where the mass 
remains constant but a field-independent impurity is added to the 
Bogomolny equation.}
with inhomogeneous mass $m(x)$ of which the potential is
\begin{equation} \label{vsg}
V = 2 \frac{m(x)^2}{\beta^2} \sin^2 \frac{\beta\phi}2 
-4\frac{m'(x)}{\beta^2} \cos\frac{\beta\phi}2,
\end{equation}
where $\beta$ is a constant and the mass $m(x)$ plays the role of
the deformation parameter $g(x)$ in \eqref{vtilde}. 
The superpotential $W_0$ in \eqref{w0} is\footnote{As discussed in
section~\ref{sec2}, we do not add any field-independent term here. 
But this is not the only possibility. Another natural choice would be
$ \tilde W_0(\phi) = \frac8{\beta^2} \sin^2\frac{\beta\phi}4  
= W_0(\phi) + \frac4{\beta^2}$,
which shifts the overall energy by a constant amount $4(m_R - m_L)/\beta^2$.}
\begin{equation}
        W_0(\phi) = -\frac4{\beta^2} \cos\frac{\beta\phi}2,
\end{equation}
which supports infinitely many extrema at 
\begin{equation} \label{2np}
        \phi_n \equiv \frac{2n\pi}\beta,\qquad  n \in \mathbb{Z},
\end{equation}
with
\begin{equation}
        W_0(\phi_n) = (-1)^{n+1} \frac4{\beta^2}.
\end{equation}

The Bogomolny equation \eqref{veq2} becomes
\begin{equation}
\frac{d\phi}{dx} = \frac{2m(x)}\beta \sin\frac{\beta\phi}2.
\end{equation}
The constant configurations \eqref{2np} trivially satisfy the Bogomolny 
equation and have energies 
\begin{equation} \label{econst}
        E_{\phi_n} = (-1)^{n+1} \frac4{\beta^2} (m_R - m_L),
\end{equation}
where we have introduced $m_R = m(\infty)$ and $m_L = m(-\infty)$.
The vacuum energy is the minimum of them,
\begin{equation} \label{evac}
        E_\textrm{vac} = - \frac4{\beta^2} |m_R - m_L|,
\end{equation}
which is negative unless $m_L = m_R$, and  corresponds to the half of the 
constant configurations in \eqref{2np}, 
\begin{equation} \label{shvac}
\phi(x) = \phi_{2n+\delta},\qquad (n \in \mathbb{Z}),
\end{equation}
where $\delta = \theta(m_L-m_R)$.
Therefore, the degeneracy of the constant configurations \eqref{2np} in
the original homogeneous theory is lifted unless $m_L = m_R$. 
This is of course because the second term in the potential 
\eqref{vtilde} is periodic in $\frac{4\pi}\beta$, which is twice the
period of undeformed theory.
These vacua preserve supersymmetry of the theory even though the 
vacuum energy \eqref{evac} is not zero.
The other half of constant solutions $\phi(x) = \phi_{2n+1-\delta}$
have positive energy $E_1 \equiv |E_\textrm{vac}|$. 

Now we look for other solutions of \eqref{veq2} which are not constant.
Let $\mu(x)$ be an integral of the mass function $m(x)$,
\begin{equation} \label{mux}
        \mu(x) = \int_0^x m(x') dx'.
\end{equation}
Then \eqref{veq2} can easily be integrated to
\begin{equation} \label{sgsol}
      \phi(x) = \frac4\beta \tan^{-1}[c\,e^{\mu(x)}]
                       \pmod{\tfrac{4\pi}\beta},
\end{equation}
where $c$ is an integration constant. In homogeneous
theories, one need not consider $c<0$ case since the period is 
$\frac{2\pi}\beta$. In the present case, however, we should also take $c<0$ 
case into account because, for example, the region $(-\frac{2\pi}\beta, 0]$
of $\phi$ cannot be identified with the region $(0, \frac{2\pi}\beta]$.
For convenience, we assume that $\phi$ is in the range 
$(-\frac{2\pi}\beta,\frac{2\pi}\beta]$, unless otherwise stated.

To understand the nature of solutions, we need to study 
the asymptotic behavior of $\mu(x)$,
\begin{equation} \label{muxa}
        \mu(x) \rightarrow m_{R/L}x + \cdots, \qquad x \rightarrow \pm \infty.
\end{equation}
If $m_L > 0$ and $m_R > 0$, the solution \eqref{sgsol} goes asymptotically to
\begin{equation} \label{phiasymp}
\phi(x) \rightarrow \begin{cases} \begin{array}{ll}
        0, & x\rightarrow -\infty, \\
        \varepsilon_c\frac{2\pi}\beta, & x\rightarrow \infty,
        \end{array} \end{cases}
\end{equation}
where $\varepsilon_c=\textrm{sgn}(c)$.
Thus the solution represents a kink/antikink for positive/negative $c$,
respectively. Depending on the inhomogeneity, however, the shape of the
solution can be very different from the usual (anti)kink as we will see below.
Changing the constant $c$ corresponds to changing the location of the 
(anti)kink as in the homogeneous theory. 
Recalling that there is no translation symmetry in the system, 
it is intriguing that there is still a position-like modulus parameter even 
for any inhomogeneous function $m(x)$. It is also worthwhile to note that all 
solutions asymptote to $\phi=0$ as $x \rightarrow -\infty$. As $x$ increases
from the left, $\phi$ either goes up or goes down depending on the sign 
of $c$. In other words, there is no (anti)kink starting at 
$\phi=\frac{2\pi}\beta$. 
These are missing in the supersymmetric solutions of the theory
since half of supersymmetry is broken here compared to the homogeneous theory. 

\begin{figure}[ht!]
\centering
\subfigure[ ]
{\label{fig:sg1}
\includegraphics[width=7.8cm]{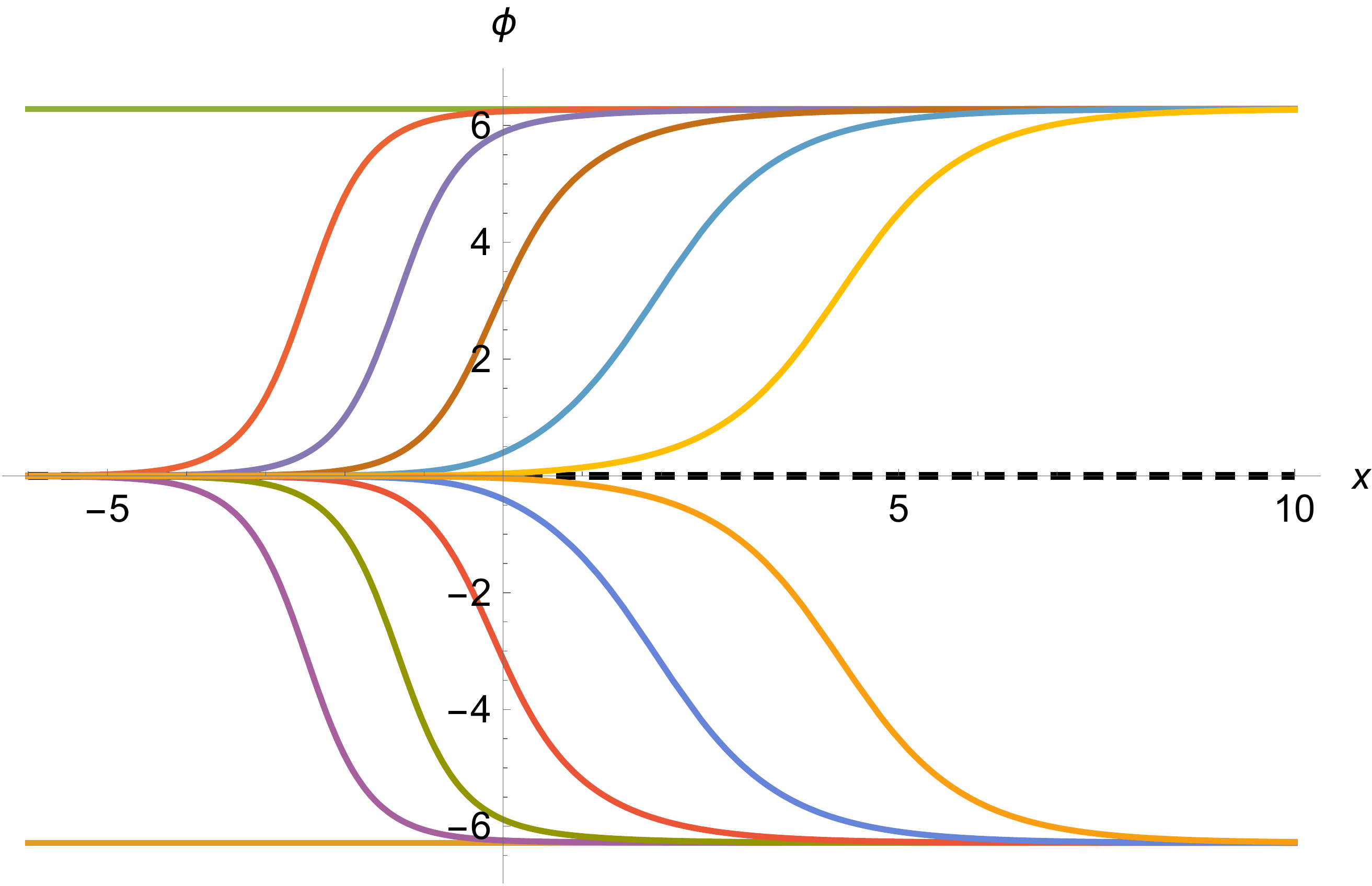}  }
\hskip0cm
\subfigure[ ]
{\label{fig:sg2}
\includegraphics[width=7.8cm]{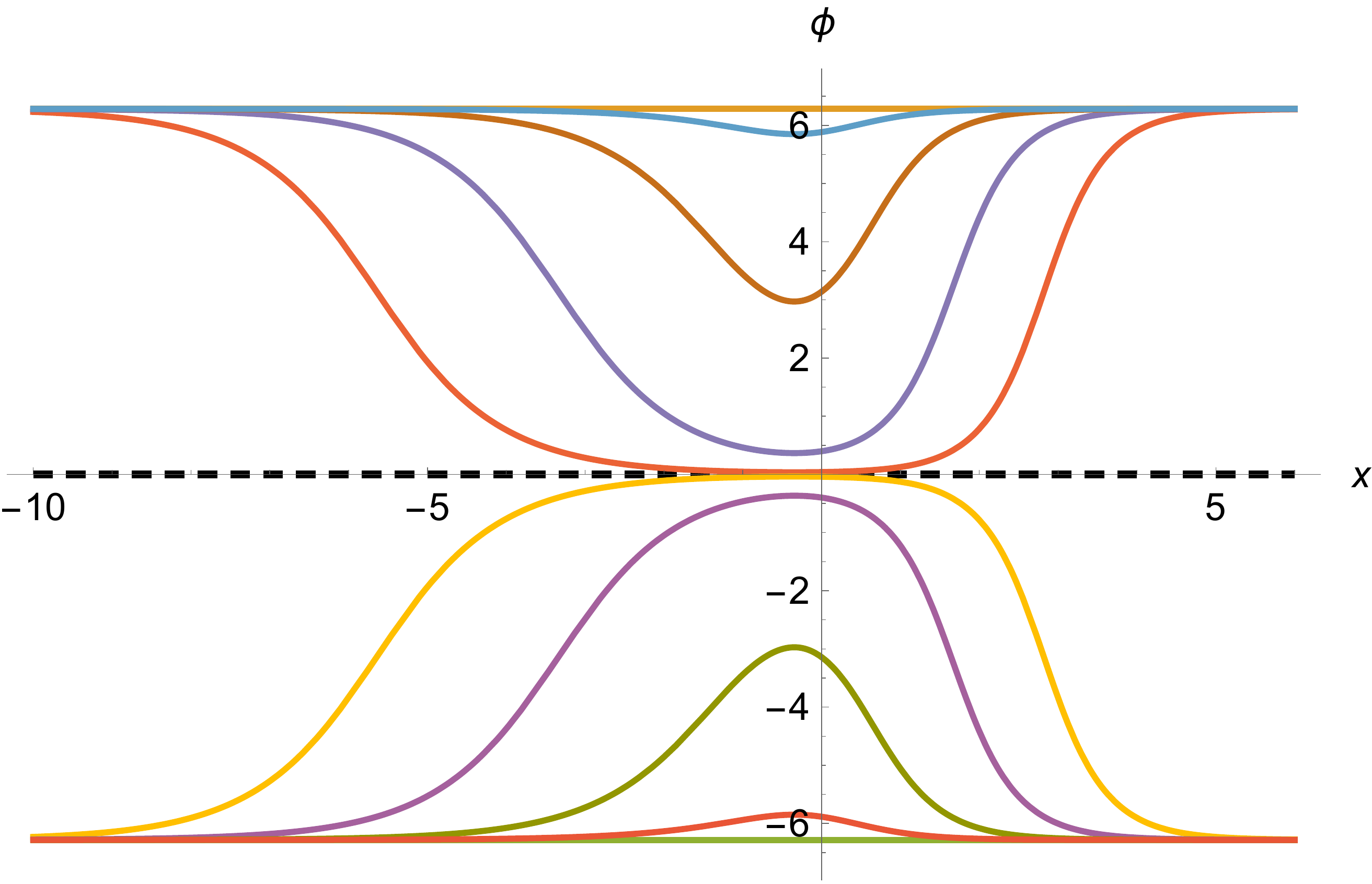}  }
\hskip0.2cm
          
\caption{Supersymmetric solutions of deformed SSG
theory for $m(x)$ in \eqref{mtanh} with $\beta=a=1$. 
(a) $m_L=2$ and $m_R=1$. The vacuum is the dashed straight line 
$\phi=0$ (mod $4\pi$) with energy 
$E=-4$. The other constant solution $\phi=2\pi$ (mod $4\pi$)
has energy $E=4$. Non-constant (anti)kink-like solutions are for 
$c=100,\ 10,\ 1,\ 0.1,\ 0.01,\ -0.01,\ -0.1,\ -1,\ -10$, and $-100$, 
respectively from the top. 
They all have energy $E_\textrm{nc}=2$.
(b) $m_L=-1$ and $m_R=2$. The vacuum is the dashed straight line 
$\phi=0$ (mod $4\pi$) with
energy $E=-12$. The other constant solution $\phi=2\pi$ (mod $4\pi$) has
energy $E=12$. Nonconstant solutions are for $c=10,\ 1,\ 0.1,\ 0.01,\
-0.01,\ -0.1,\ -1$ and $-10$, respectively from the top. 
They all have energy $E_\textrm{nc}=12$. }
\end{figure}

In Fig.~\ref{fig:sg1}, we illustrate the solutions with
\begin{equation} \label{mtanh}
m(x) = \frac12(m_R + m_L)
        + \frac12(m_R - m_L) \tanh a x, \qquad (a >0).
\end{equation}
In the figure, we set $m_L=2 > m_R=1$ so that the vacuum solution is 
$\phi(x)=0$. We can clearly see that all kinks and antikinks start at 
$\phi=0$ for $x=-\infty$. Note that the width of the (anti)kink 
solutions in the left side is thinner than that in the right side.

The energy of the solution \eqref{sgsol} denoted by $E_\textrm{nc}$ is 
given by
\begin{align} \label{enc}
E_\textrm{nc} 
&= m(\infty) W_0(\phi(\infty)) - m(-\infty) W_0(\phi(-\infty)) \notag \\
  &= \frac4{\beta^2} (|m_L| + |m_R|),
\end{align}
which is positive and independent of the integration constant $c$ as it 
should be. In this expression, we take absolute values of $m_L$ and $m_R$
which are redundant in this case but turn out to be needed in
other cases as seen below.

If both $m_L$ and $m_R$ are negative, the Bogomolny equation reduces
to the previous case by the simple parity operation 
$x \rightarrow -x$. Then the solution will be an antikink/kink for
positive/negative $c$, but now the solution starts at 
$\phi(-\infty)=\varepsilon_c\frac{2\pi}\beta$ and goes to zero as 
$x \rightarrow \infty$. The energy $E_\textrm{nc}$ would again
be given by \eqref{enc}.

When the signs of $m_L$ and $m_R$ are different from each other, the solutions
belong to a different topological sector because $\mu(x)$ in \eqref{muxa}
goes asymptotically to the same values at both spatial infinities. 
If $m_L <0$ and $m_R > 0$, $e^{\mu(x)} \rightarrow \infty$ 
as $x \rightarrow \pm\infty$ and hence
\begin{equation}
\phi(x) \rightarrow \varepsilon_c \frac{2\pi}\beta, \qquad
x \rightarrow \pm\infty.
\end{equation}
The energy $E_\textrm{nc}$ of the solution is again given by \eqref{enc}.
Finally, in the case that $m_L >0$ and $m_R < 0$, we have
\begin{equation}
\phi(x) \rightarrow 0, \qquad x \rightarrow \pm\infty,
\end{equation}
with the same energy expression \eqref{enc}. Note that in these two cases,
$\phi(x)$ asymptotes to the constant solution which is not the vacuum.

Combining with the energies for constant solutions \eqref{econst}, we have 
a rather rich BPS energy spectrum of supersymmetric solutions which depends 
on the asymptotic values of $m(x)$. If $m_R = m_L \equiv m_0$, all
constant configurations \eqref{2np} are degenerate with vanishing energy.
There are also non-constant solutions which interpolate between two adjacent
constant values of $\phi_n$ in \eqref{2np} with energy 
$E_\textrm{nc} = \frac8{\beta^2} |m_0|$.
In the homogeneous case that $m(x)=m_0$ everywhere, these are nothing 
but the kink and the antikink solutions with $\mu = m_0x$ in \eqref{sgsol}.
In the present case, however, there are two differences from the usual 
homogeneous case: half of the (anti)kink is missing as explained
above, and the shape of the solutions can be very complicated in general, 
depending on the functional form of $m(x)$. 
Nevertheless, the energy does not change under arbitrary inhomogeneous 
deformations but is solely determined by the asymptotic values of $m(x)$.

If $m_R \neq m_L$, we have two energy levels $E_\textrm{vac}$ and
$E_1 = |E_\textrm{vac}|$ for constant solutions \eqref{2np}, as described 
above. In addition, we have non-constant solutions with energy $E_\textrm{nc}$
in \eqref{enc}. For the case that $m_L$ and $m_R$ have the same sign,
$E_\textrm{nc}$ is higher than those of constant solutions. Thus we have
three discrete energy levels for supersymmetric solutions
\begin{equation}
E_\textrm{vac} < 0 < E_1 = |E_\textrm{vac}| < E_\textrm{nc}.
\end{equation}
On the other hand, if the signs of $m_L$ and $m_R$ are different, 
the energy $E_\textrm{nc}$ of non-constant solutions obtained in \eqref{enc}
is the same as $E_1$. In other words, they become degenerate and we have 
only two energy levels,
\begin{equation}
E_\textrm{vac} < 0 < E_1 = |E_\textrm{vac}| = E_\textrm{nc}.
\end{equation}
In fact, the constant solution $\phi(x) = \phi_{2n+1-\delta}$ 
is a special limit of the non-constant solution \eqref{sgsol} obtained
by taking either $c=0$ or $c=\infty$ limit. So they form a continuously 
deformable one-parameter family of degenerate
solutions where the constant solution lies in the middle of the
family. Note that there is no (anti)kink-like solutions in this case.
Also the integration constant $c$ is no longer directly interpreted as 
position of the solution. It determines the height of the solution 
rather than the location. For sufficiently large $c$, however, the 
solution may be interpreted as a threshold bound state of a kink-antikink pair
and then $c$ would determine the separation distance of them. This can
be seen in Fig.~\ref{fig:sg2} where solutions are depicted for
$m_L=-1$ and $m_R=2$. In the figure, when $|c| \ll 1$, we can clearly identify
a kink-antikink pair in the solution. As $|c|$ increases the kink and the
antikink start loosing their identity and eventually become the non-vacuum
constant configuration. It should be noted that these are all static 
supersymmetric configurations, in contrast to the standard sine-Gordon theory
where the kink-antikink pair, or the breather solution is time-dependent.
Of course, in the inhomogeneous theory this is possible due to the extra
term $\frac{\partial W}{\partial x}$ in the potential so that the interaction
between the kink and the antikink is cancelled out.

We would like to note that, of the two constant solutions, 
there is no further solution around the true vacuum, while there are
more solutions around the non-vacuum constant solution when $m_L\,m_R<0$
and they together form a continuously deformable family of degenerate solutions.
This is not just a specific feature of the sine-Gordon theory. 
Indeed, the same phenomenon occurs also in the deformed $\phi^4$ theory which 
is briefly discussed in the Appendix. Moreover, as we see in the next section,
there is a similar phenomenon in $\phi^6$ theory where the vacuum 
expectation value is inhomogeneously deformed, which is completely 
different from the inhomogeneity discussed here. Apparently, this comes from
the fact that the energy of static supersymmetric solutions depends only 
on the boundary values of the superpotential. Thus if there are any 
non-constant solutions with the same boundary values as those of a 
constant solution, they must be energetically degenerate. Since 
non-constant solutions should have a free 
parameter as argued above, it could potentially be connected to the 
constant solution, although not always guaranteed to be so. 
It would be interesting to figure out what class of inhomogeneous 
theories would have this feature.

\begin{figure}[bt!] 
\centering
\includegraphics[width=7.8cm]{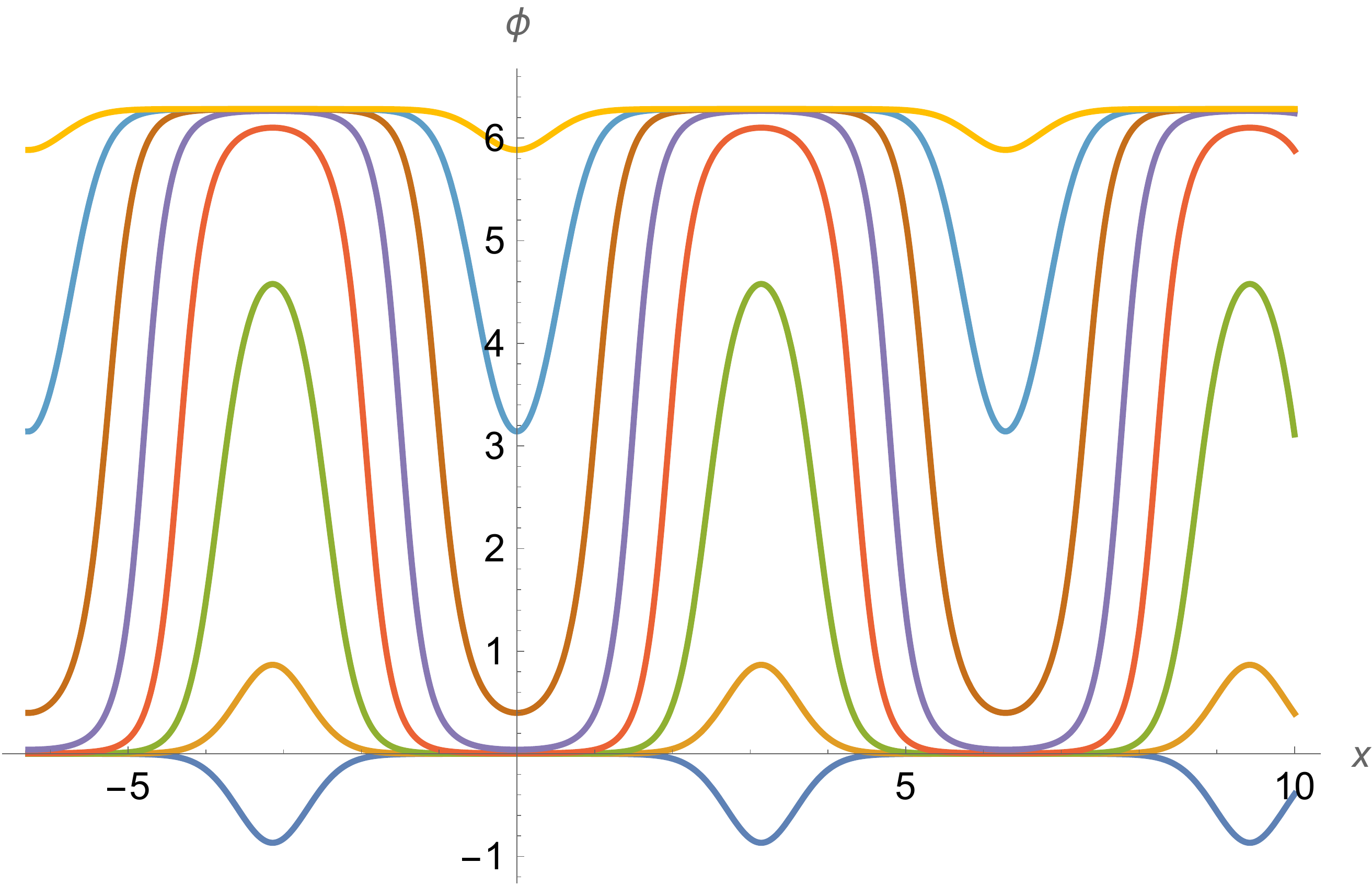}
\caption{Supersymmetric solutions of deformed SSG 
theory for $m(x)=5\sin x$
with $\beta=1$. From the top, $c=10,\ 1,\ 0.1,\ 10^{-2},\ 10^{-3},\ 10^{-4}$, 
and $-10^{-4}$, respectively.}
\label{figsg3}
\end{figure}

For different choice of $m(x)$, the profile of solutions can be arbitrarily
complicated. The (anti)kink solution or kink-antikink pair solution
described above can actually consist of many kinks and antikinks. As a
simple illustration, we depict the solution for 
\begin{equation}
m(x) = 5\sin x
\end{equation}
with $\beta=1$ for various integration constant $c$ in Fig.~\ref{figsg3}. 
Since it is periodic, it forms an infinite array of kink-antikink pairs. 
Furthermore, it is smoothly deformed to constant solutions $\phi=0$ or
$\phi=2\pi$, as $c$ goes to $0$ or $\infty$. 
The periodic inhomogeneous deformation would be relevant when the system 
is put in a circle. This example demonstrates that, in this case,
there is no longer a potential barrier between the constant solutions 
in the space of solutions. Thus there is a flat direction at the minima
of the potential in the space of solutions and they are connected by 
one-parameter family of supersymmetric solutions. The same phenomenon
would happen in other inhomogeneous theories.

We conclude this section by considering the sharp interface limit for 
the mass function \eqref{mtanh}.  If $a \rightarrow \infty$, it reduces to
\begin{equation}
        m(x) = m_L \theta(-x) + m_R \theta(x),
\end{equation}
and the potential \eqref{vsg} becomes
\begin{equation}
V = \theta(-x) 2\frac{m_L^2}{\beta^2} \sin^2 \frac{\beta\phi}2
  + \theta(x) 2\frac{m_R^2}{\beta^2} \sin^2 \frac{\beta\phi}2
  - \delta(x) \frac{4 (m_R - m_L)}{\beta^2} \cos \frac{\beta\phi}2.
\end{equation}
Then the last term of the potential proportional to the delta function 
is a pointlike impurity at the origin. Since the original
potential depends only on the squares of the masses, there are four 
possibilities in the impurity strength for given $m_L^2$ and $m_R^2$,
i.e., $m_L-m_R$ can be identified as $\pm(|m_L| + |m_R|)$ or 
$\pm(|m_L|-|m_R|)$ depending on the signs of $m_L$ and $m_R$.
The resulting quantum theory could be further studied using various
techniques in quantum field theory, which is beyond the scope of this paper.

\section{Inhomogeneous Deformation of the Vacuum Expectation Value 
in $\phi^6$ Theory}
\label{sec4}

Let us consider a quartic superpotential,
\begin{equation}
        \frac\lambda4 (\phi^4 - 2 w \phi^2).
\end{equation}
If the parameters $\lambda$ and $w$ are constants, it results in the 
sextic potential
\begin{equation}
\frac{\lambda^2}2 \phi^2 ( \phi^2 - w )^2.
\end{equation}
For a positive $w$, this potential is triply degenerate
with vacua at $\phi = 0, \pm \sqrt w$. On the other hand, for negative $w$,
there is only one vacuum at the origin $\phi=0$.
If $\lambda$ is dependent on the position, we would get similar results to
the previous section. Here, we let $w$ be inhomogeneous instead,
\begin{equation}
        W(\phi,x) = \frac\lambda4 [\phi^4 - 2 w(x) \phi^2],
\end{equation}
so that the corresponding potential \eqref{vphix} becomes
\begin{equation} \label{vphi6}
        V(\phi,x) = \frac{\lambda^2}2 \phi^2 ( \phi^2 - w(x) )^2
- \frac\lambda2 w'(x) \phi^2.
\end{equation}
\begin{figure}[ht!]
\centering
   \includegraphics[width = 9.5cm]{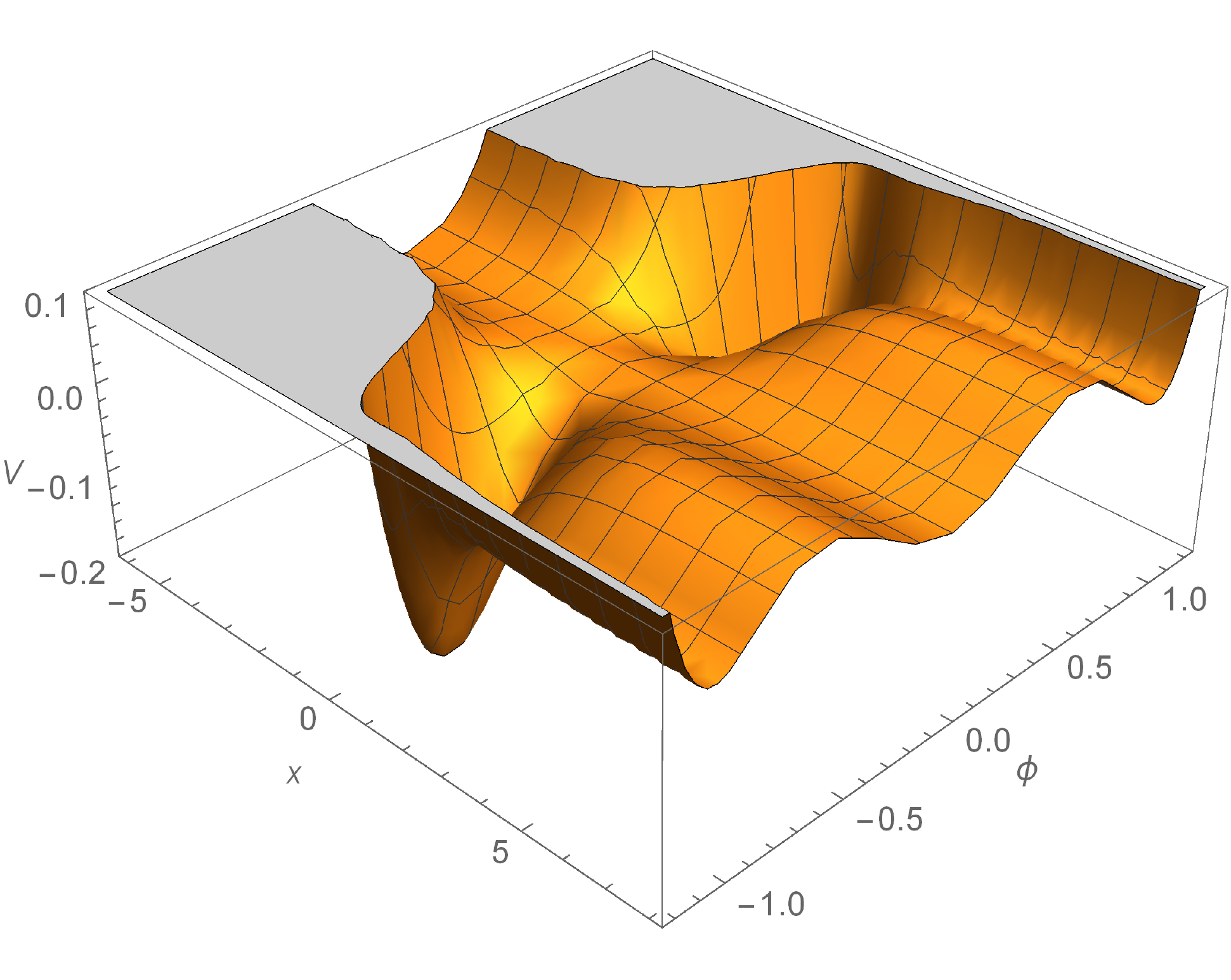}
    \caption{An example of the inhomogeneous potential $ V(\phi, x)$ in 
	\eqref{vphi6} with $w(x)=\tanh x$ and $v=\lambda = 1$.  } \label{Vphix}
\end{figure}
We illustrate the shape of the potential in Fig.~\ref{Vphix} for 
$w(x) = \tanh x$ and $\lambda=v=1$.

It is important to note the difference from the overall rescaling considered
in the previous section. 
In the present case, the minima of the potential \eqref{vphi6}
depend on the position and even the total number of the minima can change
as shown in Fig.~\ref{Vphix}, since $w$ may not be positive definite. 
One may think that half of the system lies in some different physical
situation such as external field, or two heterogeneous systems with distinct 
potentials smoothly joined together while supersymmetry is preserved.
Then the vacuum need not just be given by a constant configuration in 
general, in contrast with the previous case where the vacuum 
configurations are constants since the overall shape of the superpotential 
is not deformed.

Supersymmetric solutions satisfy the Bogomolny equation
\begin{equation} \label{eq6}
        \phi'(x) - \lambda \phi(x) [ \phi^2(x) - w(x) ] = 0.
\end{equation}
We assume that $w(x)$ goes to some finite values $w_{L/R}$ as
$x\rightarrow \pm\infty$ which are not necessarily positive. Then possible
asymptotic values of $\phi(x)$ depend on the signs of $w_L$ and
$w_R$. If both $w_L$ and $w_R$ are negative, 
$\phi(x)$ should vanish at spatial infinities to achieve the minimum energy. 
If one or both of $w_L$ and $w_R$ are positive, $\phi$ can go to 
$\pm\sqrt{w_L}$ or $\pm\sqrt{w_R}$ at spatial infinities. 
The superpotential at these values are
\begin{align}
        W(0) &= 0, \nonumber \\
        W(\pm \sqrt{w_R}) &= -\frac\lambda4 w_R^2, \nonumber \\
        W(\pm \sqrt{w_L}) &= -\frac\lambda4 w_L^2.
\end{align}

Now let us solve the Bogomolny equation \eqref{eq6}. An obvious solution is
\begin{equation}
        \phi(x) = 0,
\end{equation}
of which the energy vanishes. This constant solution is possible since
$\phi(x) = 0$ is still an extremum of the potential irrespective of the 
function $w(x)$.
There exist other solutions. In fact, the Bogomolny equation \eqref{eq6} is 
integrable for arbitrary $w(x)$. The general solution is given by
\begin{equation} \label{sol6}
        \phi^2(x) = \frac{e^{-2\lambda \Omega(x)}}{2\lambda(c-\xi(x))},
\end{equation}
where $c$ is an integration constant and
\begin{align} \label{ox}
        \Omega(x) &= \int w(x) dx, \nonumber \\
        \xi(x) &= \int e^{-2\lambda \Omega(x)} dx.
\end{align}
Then $\Omega(x) \sim w_{R/L} x$ as $x \rightarrow \pm \infty$.
Since $\phi^2(x)$ is positive, $\xi(x)$ in the denominator of \eqref{sol6}
should be bounded from above by the integration constant $c$. From \eqref{ox}, 
$\xi(x)$ is monotonically increasing and then the asymptotic values of $w(x)$ 
is restricted to be $w_R \ge 0$ for the solution not to be divergent, 
while $w_L$ can take any value.

The energy of the solution is calculated from the asymptotic values of 
$\phi(x)$ at spatial infinities $x \rightarrow \pm\infty$.
Since the theory is symmetric under the sign flip $\phi \rightarrow -\phi$, 
it is enough to take into account $\phi(\infty) \ge 0$ case only.
Let us first consider $x \rightarrow -\infty$ limit in which $\phi(-\infty)$ 
is determined by $w_L$. If $w_L > 0$, both the numerator and the denominator 
of \eqref{sol6} diverges and the limit can be calculated with the help of
L'H\^opital's rule,
\begin{equation}
\phi^2(x) \xrightarrow{x\rightarrow -\infty} 
\frac{-2\lambda w(x) e^{-2\lambda \Omega(x)}}%
       {- 2\lambda e^{-2\lambda \Omega(x)}} \longrightarrow w_L.
\end{equation}
For $w_L \le 0$, it is easy to see that the numerator of \eqref{sol6} vanishes
while the denominator does not in $x \rightarrow -\infty$ limit, i.e.,
$\phi(-\infty)=0$. 

Next, we consider $x \rightarrow \infty$ limit.
From \eqref{sol6}, we see that $\phi(\infty)$ depends on 
the value of the denominator. If $\xi(\infty)$ is strictly less than $c$, 
$\phi^2(x) \sim e^{-2 \lambda w_R x}$ as $x \rightarrow \infty$ and hence 
the solution vanishes $x \rightarrow \infty$.
If $\xi(\infty)=c$, the denominator in \eqref{sol6} vanishes asymptotically.
In this case, the solution \eqref{sol6} may be written in
a more explicit form by absorbing the integration constant $c$ as
\begin{equation} \label{sol62}
\phi^2(x) = \frac{e^{-2\lambda \Omega(x)}}%
        {2\lambda \int_x^\infty e^{-2\lambda \Omega(x')} dx'}.
\end{equation}
In $x \rightarrow \infty$ limit, we again use L'H\^opital's rule to
get $ \phi^2(\infty) = w_R$.
In other words, $\phi(\infty)$ vanishes for all integration constant $c$ 
except one specific value for which $\phi(\infty)=\sqrt{w_R}$. 
It is amusing to see how the solution with 
different boundary condition arises naturally for a special value of 
the integration constant.

We have now four different asymptotic values at spatial infinities, 
depending on the integration constant and the sign of $w_L$. For each case, 
the corresponding energy can be calculated as the difference of the 
superpotential as in Table~1.

\begin{table}
\centering
\begin{tabular}{ c | c | c | c | c }
\hline
$w_L$ & $c$ &  $\phi^2(-\infty)$ & $\phi^2(\infty)$ & $E$ \\
\hline
\multirow{2}{3.5em}{$w_L>0$} & $c > \xi(\infty)$ & $w_L$ & 0 & $\frac\lambda4 w_L^2$ \\
    & $c = \xi(\infty)$ & $w_L$ & $w_R $ & $\frac\lambda4 (w_L^2 - w_R^2)$ \\
\hline
\multirow{2}{3.5em}{$w_L \le 0$} & $c > \xi(\infty)$ & 0 & 0 & 0 \\
    & $c = \xi(\infty)$ & 0 & $w_R$ & $-\frac\lambda4 w_R^2$ \\
\hline
\end{tabular}
\caption{Asymptotic values and energies of the non-constant solution 
\eqref{sol6} in deformed $\phi^6$ theory. In addition, there is a constant
solution $\phi=0$ with $E=0$.}
\end{table}

Let us now summarize the supersymmetric solutions of the theory. 
There is always a trivial solution $\phi(x) = 0$ with vanishing energy. 
If $w_L>0$, we have a unique solution\footnote{To be
more precise, there is also another solution trivially obtained by the 
sign flip $\phi \rightarrow -\phi$ which interpolates between
$\phi(-\infty)=-\sqrt{w_L}$ and $\phi(\infty)=-\sqrt{w_R}$. 
The flipped solution is always assumed to exist below and will not
be discussed explicitly, unless otherwise stated.}
\eqref{sol62} interpolating between $\phi(-\infty)=\sqrt{w_L}$ and 
$\phi(\infty)=\sqrt{w_R}$. The energy of these solutions is
$E= \frac\lambda4 (w_L^2-w_R^2)$ which can either be positive or negative
depending on the value of $w_L$ and $w_R$. If $w_L < w_R$, the energy
becomes lower than that of the trivial $\phi=0$ solution and we have
two degenerate vacua (including the sign-flipped solution) with negative 
energy. If $w_L=w_R$ we would have triply degenerate vacua with vanishing 
energy. There are also a one-parameter family of degenerate solutions 
with higher energy $E = \frac\lambda4 w_L^2$.

If $w_L \le 0$, we have again a unique solution \eqref{sol62}
which now interpolates between $\phi(-\infty)=0$ and $\phi(\infty)=\sqrt{w_R}$
with negative energy $E = -\frac14 \lambda w_R^2$. All the other
solutions have vanishing energy irrespective of the value of $c$, since
the energy depends only on the boundary values of $\phi$. They 
form a one-parameter family of degenerate solutions which includes the 
constant solution $\phi(x)=0$ as a special $c \rightarrow 0$ limit, 
similar to what we have seen in the previous section.

Note that, even though we have a constant solution $\phi(x)=0$ with vanishing
energy, this is not the vacuum solution of minimum energy except
$w_L > w_R$ case. The true vacuum is not constant but interpolates between two 
different values of the minima with negative energy: 
either $\phi(-\infty)=\sqrt{w_L}$ and $\phi(\infty)=\sqrt{v_R}$ (for 
$w_R > w_L \ge 0$), or $\phi(-\infty)=0$ and $\phi(\infty)=\sqrt{w_R}$ 
(for $w_L < 0 < w_R$).

As an explicit example, let us consider a specific spatial dependence as before,
\begin{equation} \label{wx}
w(x) = \frac12(w_R + w_L) + \frac12(w_R - w_L) \tanh a x, \qquad (a>0).
\end{equation}
Then we can express the solution \eqref{sol6} in terms of a hypergeometric 
function which is rather complicated and it is not much illuminating to show 
it here. Instead, we choose some simple $w_L$'s for fixed $w_R$ and see 
explicitly how the solution changes as we vary $w_L$.
\begin{itemize}
\item 
If $w_L=w_R$, $w(x)$ is constant and we have a usual homogeneous 
$\phi^6$ theory. Then the solution reduces to~\cite{Lohe:1979mh}
\begin{equation} \label{antikinkh}
\phi(x) = \pm\left(\frac{w_R}{1 + c\, e^{2 \lambda w_R x}}\right)^{1/2},
\end{equation}
which are antikink/kink solutions interpolating between 
$\phi(-\infty)=\pm\sqrt{w_L}$ respectively, and $\phi(\infty)=0$ if $c>0$. 
When $c=0$, the solutions become constant solutions $\phi(x)= \pm\sqrt{w_R}$ 
which are just the vacua degenerate with $\phi(x)=0$ configuration. 
Note that, in homogeneous case, the supersymmetry is enhanced and
we have other supersymmetric solutions which are kinks/antikinks 
interpolating 0 and $\pm\sqrt{w_R}$, respectively. They are, however, 
not solutions of \eqref{eq6} but solutions of the equation with positive 
sign in front of $\lambda$ in \eqref{eq6}.

\item If $w_L = 2 w_R$ and $a = \lambda w_R$, the solution is expressed
in terms of elementary functions,
\begin{equation}
\phi(x) = \pm\left(\frac{2w_R(1+e^{2\lambda w_R x})}%
  {1 + 2e^{2\lambda w_R x} + c\, e^{4 \lambda w_R x}}\right)^{1/2},
\end{equation}
which interpolates between $\pm\sqrt{w_L}(=\pm\sqrt{2w_R})$ and 0 for $c>0$,
respectively, and $\pm\sqrt{w_L}$ and $\pm\sqrt{w_R}$ for $c=0$, respectively. 
Since $w_L > w_R$, these solutions have positive energies and
the vacuum solution is $\phi(x)=0$. 
We draw the solutions for various $c$ in Fig.~\ref{fig:w1} with 
$a=\lambda=w_R=1$. The vacuum solution is depicted with a dashed 
line. For large $c$, the (anti)kink lies in the left side. As $c$ gets smaller,
it moves to the right side but the height of the solution
changes to $\sqrt{w_R}$ so that it looks like double (anti)kinks separated
with the left (anti)kink being fixed around $x=0$.
When $c = 0$, the boundary value at $x=\infty$ changes
from $\phi=0$ to $\phi=\pm\sqrt{w_R}$ which reduces the energy
from $E=\lambda w_R^2$ to $E=\frac34 w_R^2$.
\begin{figure}[htbp]   
\centering
\subfigure[]
{ \label{fig:w1}
   \includegraphics[width=0.4\textwidth]{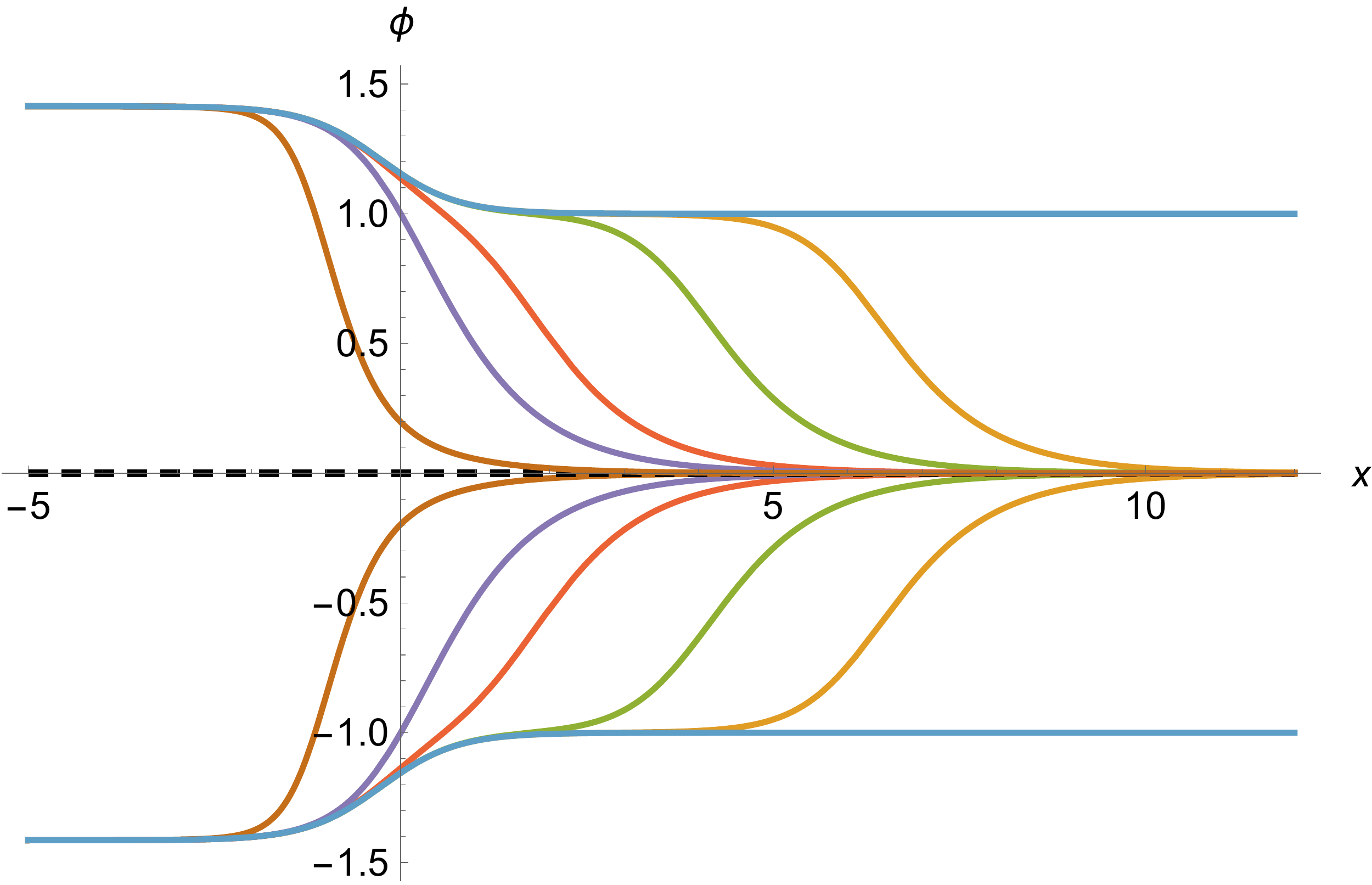}
}
\subfigure[]
{ \label{fig:w2}
   \includegraphics[width=0.4\textwidth]{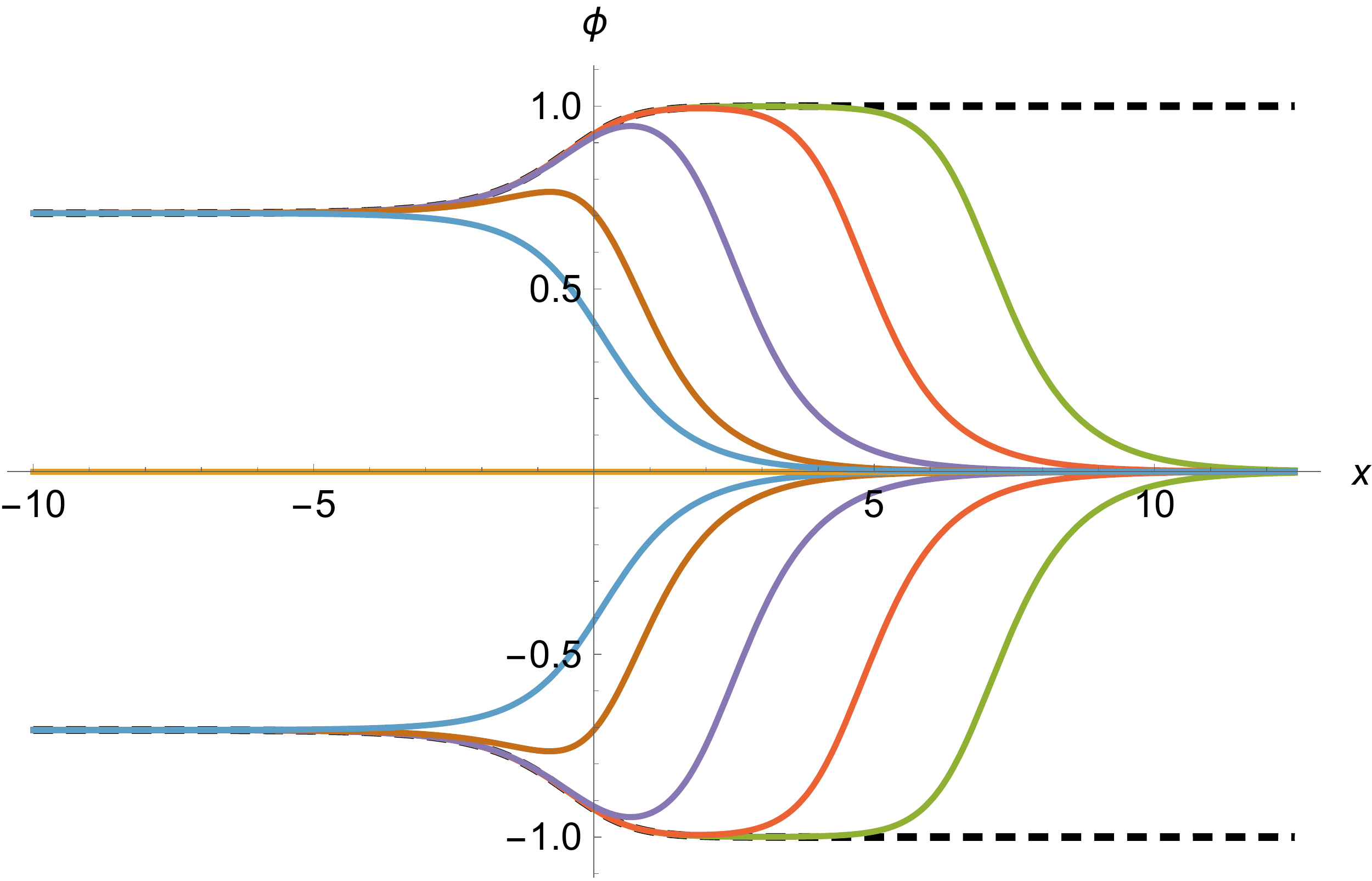}
}
\subfigure[]
{ \label{fig:w3}
   \includegraphics[width=0.4\textwidth]{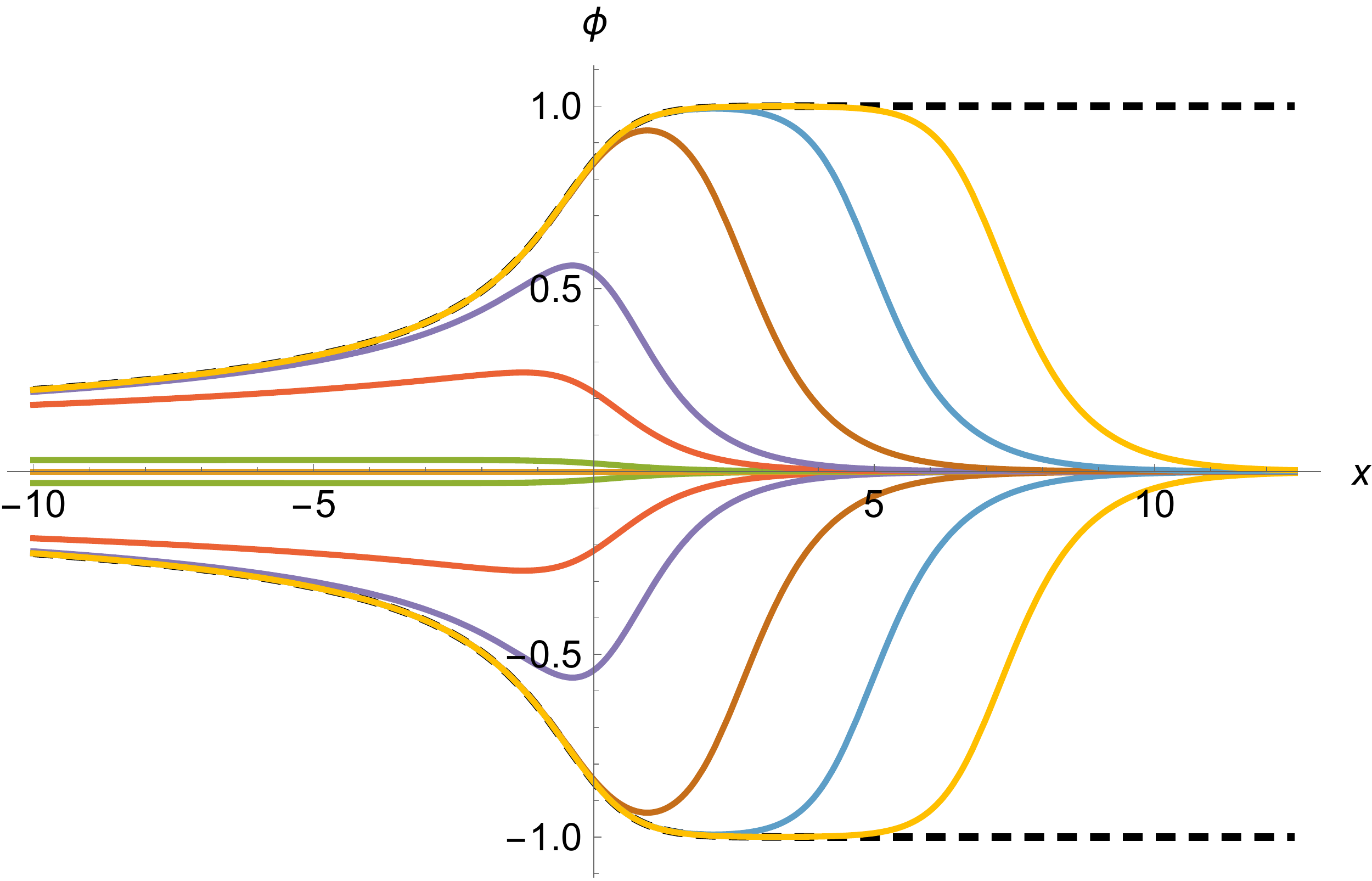}
}
\subfigure[]
{ \label{fig:w4}
   \includegraphics[width=0.4\textwidth]{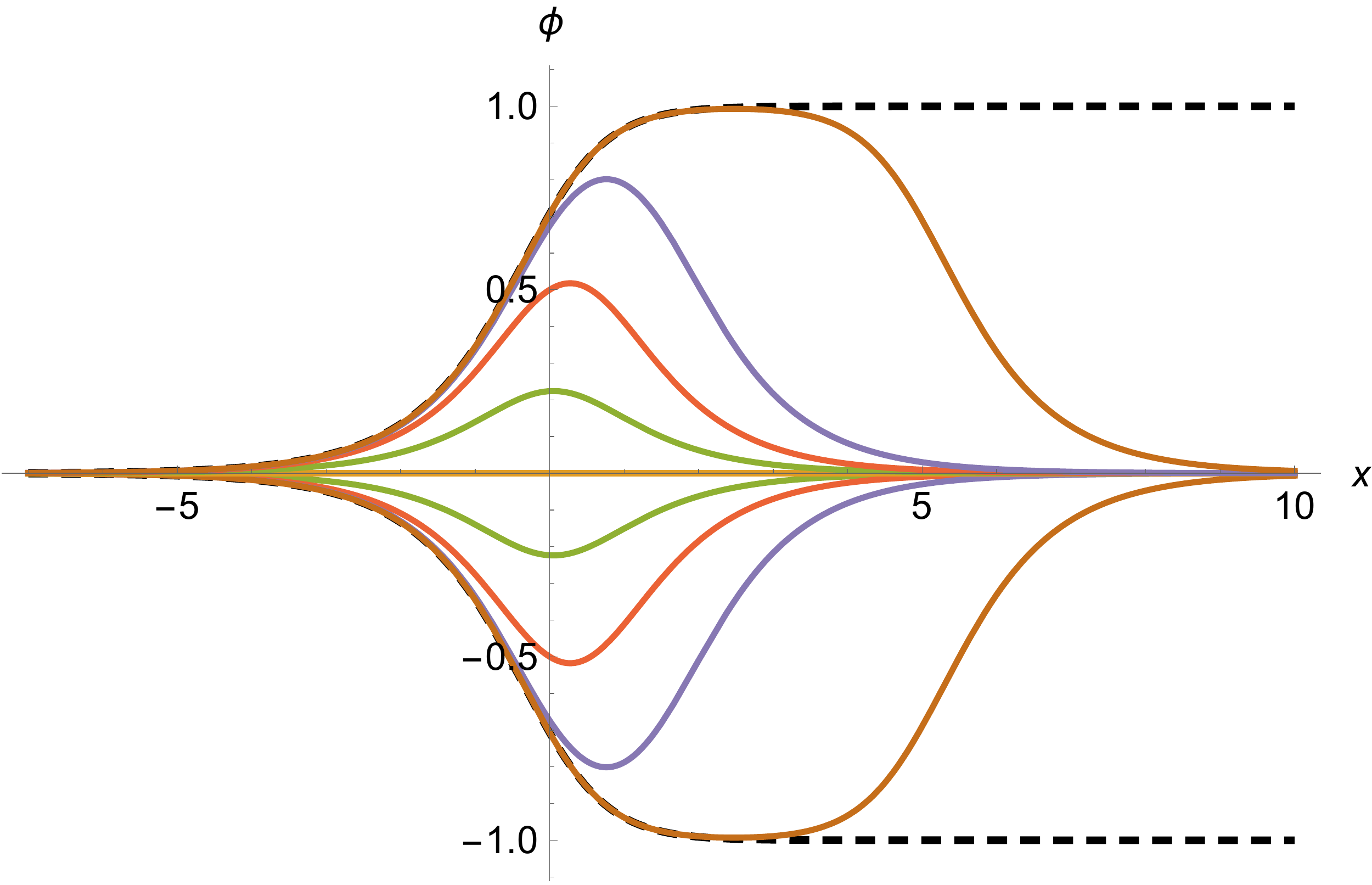}
}
\caption{Supersymmetric solutions of the deformed $\phi^6$ theory for 
$w(x)$ given in \eqref{wx} with $a=\lambda=w_R=1$. Solutions in $\phi<0$ 
region are obtained by $\phi \rightarrow -\phi$. 
In each figure, vacuum solutions are depicted with dashed lines.
(a) Plots for $w_L=2$. The topmost curve interpolates between $\sqrt2$ and 1
and has energy $E=\frac34$. The other solutions interpolating between $\sqrt2$ 
and 0 are for $c=100,\ 1,\ 0.1,\ 10^{-3}$, and $10^{-5}$, respectively 
from the left. The energy of these solutions are all $E=1$.
(b) Plots for $w_L=\frac12$. The dashed curves interpolating between 
$\pm\frac1{\sqrt2}$ and 1 are vacuum solutions with negative energy 
$E=-\frac3{16}$. The other solutions interpolating between $\frac1{\sqrt2}$ 
and 0 are for $c=1,\ -1,\ -\sqrt2+10^{-2},\ -\sqrt2+10^{-4}$, and 
$-\sqrt2+10^{-6}$, respectively from the left. The energy of these solutions 
are all $E=\frac1{16}$.
(c) Plots for $w_L=0$. The dashed curves interpolating between 0 and $\pm 1$
are vacuum solutions with negative energy $E=-\frac14$. The other 
solutions all vanish as $x\rightarrow \pm\infty$ and are for $c=10^{-6},\ 
10^{-4},\ 10^{-2},\ 1,\ 10,\ 10^3$ and $\infty$, respectively 
from the top. The energy of these solutions are all zero.
(d) Plots for $w_L=-1$. The vacuum solutions are dashed curves 
interpolating between 0 and $\pm 1$ with energy $E=-\frac14$. 
The other solutions 
vanishing at both spatial infinities are for $c=1.0001,\ 1.1,\ 2,\ 10$, and 
$\infty$, respectively from the top. The energies of these solutions are all 
zero.
}
\label{fig:phi66}
\end{figure}

\item If $w_L = \frac{w_R}2$ and $a = \lambda w_R$, we have
\begin{equation}
\phi(x) = \pm\left(\frac{w_R}2 \frac1{1 + e^{2\lambda w_R x}%
            +c\, e^{3 \lambda w_R x/2}\sqrt{\cosh\lambda w_R x}}\right)^{1/2},
\end{equation}
where $c>-\sqrt2$ corresponds to the solution interpolating between 
$\pm\sqrt{w_L}(=\pm\sqrt{\frac{w_R}2})$ and 0
with energy $E=\frac\lambda4 w_L^2 = \frac\lambda{16}w_R^2$,
while $c=-\sqrt2$ gives the vacuum solutions interpolating between 
$\pm\sqrt{w_L}$ and $\pm\sqrt{w_R}$, respectively, with negative energy 
$\frac\lambda4(w_L^2 - w_R^2) = -\frac{3\lambda}{16}w_R^2$.
We draw the solutions in Fig.~\ref{fig:w2} with $a=\lambda=w_R=1$ where
the dashed lines represent the vacuum solutions which are no longer constants.

\item If $w_L = 0$ and $a = \lambda w_R$, the solution becomes
\begin{equation}
\phi(x) = \pm\left(\frac{w_R}2 \frac{1-\tanh \lambda w_R x}%
               {c + \ln(1+e^{-2\lambda w_R x})}\right)^{1/2}.
\end{equation}
For $c>0$, the solutions vanish at both spatial infinities and hence
have vanishing energy. The constant solution $\phi(x)=0$ is obtained in
$c\rightarrow \infty$ limit. For $c=0$, they interpolate between 0 and 
$\pm\sqrt{w_R}$ with negative energy $E=-\frac\lambda4 w_R^2$ and hence 
they are  the vacuum solutions. Note that, with this form of $w(x)$, the 
potential smoothly interpolates between the unbroken phase with pure 
$\phi^6$ potential in the left side of the space and the broken phase 
with three degenerate minima at 0 and $\pm\sqrt{w_R}$ in the right side 
of the space, while keeping the supersymmetry. 
We draw the solutions in Fig.~\ref{fig:w3} with $a=\lambda=w_R=1$. 
The vacuum solutions are depicted with dashed lines.

\item Finally, we consider the case $w_L = -w_R <0$ 
and $a = \lambda w_R$, for which $w(x)$ is simply given by
\begin{equation} \label{omegaxx}
w(x) = w_R \tanh \lambda w_R x.
\end{equation}
Then in the left half of space the sextic potential has only one
minimum at the origin $\phi=0$, while, in the right half, it has three
minima at $\phi=0$ and $\pm \sqrt{w_R}$. Now the solution \eqref{sol6} becomes
\begin{equation} \label{phitanh}
\phi(x) = \pm\left(\frac{w_R}2\frac{\sech^2\lambda w_R x}%
                     {c - \tanh\lambda w_R x}\right)^{1/2},
\end{equation}
which requires $c \ge 1$. If $c>1$, these solutions vanish at both
spatial infinities and the corresponding energy is zero. The solutions
with different $c$ are all degenerate and the constant solution 
$\phi(x)=0$ is obtained in $c \rightarrow \infty$ limit.
If $c=1$, \eqref{phitanh} reduces to
\begin{equation} \label{vactanh}
\phi(x) = \pm\left[\frac{w_R}2 (1 + \tanh \lambda w_R x)\right]^{1/2},
\end{equation}
which interpolates between 0 and $\pm\sqrt{w_R}$, respectively, with 
energy $E=-\frac14 \lambda w_R^2$. 
Note that the shape of \eqref{vactanh} is identical to the kink/antikink 
solutions in the homogeneous theory which are given in \eqref{antikinkh} 
with $c=1$ and $x \rightarrow -x$. In our case, however, these configurations
are the vacuum solutions with negative energy.
The solutions are depicted in Fig.~\ref{fig:w4} with $a=\lambda=w_R=1$.
As we seen in this case and also in the $w_L=0$ case discussed above, 
the constant solution $\phi(x)=0$ and non-constant solutions around it which
have the same boundary values form a continuously deformable one-parameter
family of degenerate solutions, a phenomenon seen in the previous section.
\end{itemize}

\section{Discussions}
\label{sec5}
We studied inhomogeneous supersymmetric field theories classically 
in two dimensions where the superpotential depends explicitly on the spatial
coordinate. Though the translational invariance is broken, half of 
the supersymmetry can be preserved by adding a term in the Lagrangian
which is simply given by the spatial derivative of the superpotential.
Analyzing the superalgebra, we showed that the energy is bounded from below by
a topological charge which is not necessarily nonnegative definite.
The energy bound is saturated for supersymmetric 
configurations satisfying the first-order Bogomolny equation. 

There would be, in principle, many different types of inhomogeneity.
In this paper, we considered two types of inhomogeneity: One is
the inhomogeneous rescaling of the superpotential 
for SSG theory and the other is inhomogeneous 
deformation of the vacuum expectation value in $\phi^6$ theory. 
For these theories, we obtained the most general solutions of the Bogomolny
equation and their energies which depend only on the boundary values of the 
superpotential.
There is in general a zero mode around the non-vacuum solution which
persists even under any arbitrary inhomogeneous deformation.
Since the deformation function is arbitrary, one can consider various
specific forms to obtain interesting physical situations such as sharp
interface limit, or periodic functions.

In this paper, we considered only supersymmetric solutions of three 
classical theories in two dimensions. It would be interesting to investigate 
dynamical as well as quantum aspects of the theories in detail including 
zero mode dynamics and higher excitations. The structure of the solution
space could be much richer in inhomogeneous theories as we have seen in this 
paper. In addition, we may consider other 
kind of inhomogeneities not studied in this paper. 
An immediate possibility would be to perform deformations of 
both the rescaling and the vacuum expectation value at the same time for
theories considered here or for other theories.
One may also deform other parameters not considered here.
Another possibility is to deform the form of the potential by adding new 
terms. In this direction, impurities were studied in \cite{Adam:2019yst} 
by adding new terms to the original Lagrangian as in \eqref{adamL2}. 
Also the kinetic term can have inhomogeneity as seen in \eqref{Im2dQFT2}.

Since the superpotential $W$ is arbitrary in our formulation, one can consider
inhomogeneous deformations of any existing homogeneous theories other than 
those studied here. For example, double SG 
model~\cite{condat1983, Gani:2017yla} and models with higher-order
polynomial potentials~\cite{Khare:2014kva,Christov:2018wsa} 
would be natural candidates to extend the results of
SG theory and $\phi^6$ theory, respectively.

More generally, one may imagine two or more heterogeneous
systems smoothly joined together with half of the supersymmetry being kept. 
The $\phi^6$ theory in section~\ref{sec4} can be considered as a simple 
example of this kind since it connects two potentials with different vacuum 
structures. In this case, the vacuum configuration would be no longer 
constant in general.

It should not be difficult to consider higher-dimensional theories 
with more field contents such as gauge fields. In fact, as mentioned in 
the introduction, there are already many studies on inhomogeneous 
supersymmetric field theories in various dimensions.
However, we believe that the consequences of the inhomogeneity 
have not been fully explored. We will investigate these issues
in separate publications~\cite{KKKS}.

\section*{\bf Acknowledgement}

We would like to thank Changrim Ahn, Kyung Kiu Kim and Hanwool Song for useful
discussions. Also thanks are due to Wereszczynski for informing us
of their earlier works, especially \cite{Adam:2019djg, Manton:2019xiq}.
This work was supported by the National Research Foundation of
Korea(NRF) grant with grant number NRF-2019R1F1A1059220 (C.K.),
NRF-2019R1F1A1056815 (Y.K.), and NRF- 2020R1A2C1014371, NRF-2019R1A6A1A10073079
(O.K.). Hospitality at APCTP during the program ``String theory, gravity 
and cosmology (SGC2021)'' is kindly acknowledged (YK, OK).

\appendix

\section{\bf $\phi^4$ theory}

In this appendix, we consider inhomogeneous rescaling of 
$\phi^4$ theory with a double-well potential. It has already been investigated 
in~\cite{Adam:2019djg,Manton:2019xiq}. Since it is a prototypical
example, however, we summarize the results and discuss energy spectrum
which differs from \cite{Adam:2019djg} because the definition of the energy
is not the same as discussed in section~\ref{sec2}. The deformed potential 
reads
\begin{equation} \label{vphi4}
V(\phi,x) = \frac12 g^2(x) ( \phi^2 - v^2 )^2
+ \frac13 g'(x) (\phi^3 - 3 v^2 \phi),
\end{equation}
which corresponds to the superpotential $W_0$ given by
\begin{equation}
        W_0 = \int ( \phi^2 - v ^2 ) d\phi = \frac13 \phi^3 - v^2 \phi.
\end{equation}
Here we have dropped the trivial field-independent term
as discussed in section 2. $W_0$ involves two extrema 
$\phi=\pm v$ which trivially satisfy the Bogomolny equation \eqref{veq2} and
have the values
\begin{equation}
        W_0(\pm v) = \mp\frac23 v^3.
\end{equation}
The corresponding energies are
\begin{equation} \label{ev}
        E_{\phi=\pm v} = \mp \frac23 v^3 (g_R-g_L).
\end{equation}
The vacuum energy is then the minimum of the two,
\begin{equation} \label{phi4evac}
E_\textrm{vac} = -\frac23 v^3 |g_R - g_L| ,
\end{equation}
which is negative unless $g_L=g_R$. 

Depending on the magnitude of $\phi(x)$,
there are two types~\cite{Adam:2019djg} of nontrivial solutions of the 
Bogomolny equation \eqref{veq2}. The first type is given by
\begin{equation} \label{tanh}
        \phi(x) = -v \tanh(v G(x) - c),
\end{equation}
which satisfies $|\phi(x)|<v$. Here $c$ is an integration constant and 
\begin{equation} \label{Gx}
        G(x) = \int_0^x g(x') dx'.
\end{equation}
For the other type, $|\phi(x)| > v$ and the solution is\footnote{The 
solution \eqref{coth} becomes singular when $v G(x)-c$ vanishes at 
some $x$. It happens if $g_L$ and $g_R$ have the same sign as seen 
in \eqref{Gxa}. For the case that $g_L$ and $g_R$ have different signs, 
$vG(x) - c$ can vanish for some integration constant $c$. 
We will not consider these singular cases.}
\begin{equation} \label{coth}
        \phi(x) = -v \coth(v G(x) - c).
\end{equation}

The energy of the solutions may be obtained from the asymptotic behavior 
of $G(x)$,
\begin{equation} \label{Gxa}
        G(x) \rightarrow g_{R/L}x + \cdots, \qquad x \rightarrow \pm \infty.
\end{equation}
Then the solutions \eqref{tanh} and \eqref{coth} behave asymptotically as
\begin{equation} \label{phi4asymp}
        \phi(x) \rightarrow \begin{cases}
               \textrm{sgn}(g_L) v, \qquad x\rightarrow -\infty, \\
               -\textrm{sgn}(g_R) v, \qquad x\rightarrow \infty,
        \end{cases}
\end{equation}
and hence the corresponding energy denoted by $E_\textrm{nc}$ is computed
to be
\begin{align} \label{phi4enc}
E_\textrm{nc} 
&= g(\infty) W_0(\phi(\infty)) - g(-\infty) W_0(\phi(-\infty)) \nonumber \\
  &= \frac23 v^3 (|g_R| + |g_L|),
\end{align}
which is independent of the integration constant $c$.

As in the SSG case, we have different types of solutions and energy 
levels depending on the asymptotic values of $g(x)$. If $g_L\, g_R > 0$, 
the non-constant solutions are basically deformed kinks or antikinks
interpolating between two values $\phi = v$ and $\phi = -v$. Unlike the
SSG case, however, here we have only one type of solutions either
kinks or antikinks but not both. Also there is no coth-type solution
in this case. Since the energy $E_\textrm{nc}$ of non-constant solutions are
higher than those of constant solutions, we have three discrete energy
levels for supersymmetric solutions,
\begin{equation}
E_\textrm{vac} = E_{\phi = \varepsilon v} < 0 
< E_{\phi = -\varepsilon v} < E_\textrm{nc},
\end{equation}
where $\varepsilon \equiv \textrm{sgn}(g_R-g_L)$. 
If the signs of $g_L$ and $g_R$ are different, 
$\phi \rightarrow -\varepsilon v$ at both spatial infinities, 
which is not the vacuum value, and the corresponding energy $E_\textrm{nc}$ 
in \eqref{phi4enc} becomes identical to $E_{\phi=-\varepsilon v}$. Therefore
we have only two energy levels for supersymmetric solutions,
\begin{equation}
E_\textrm{vac} = E_{\phi = \varepsilon v} < 0 
< E_{\phi = -\varepsilon v} = E_\textrm{nc}.
\end{equation}
As in the SSG case, the non-constant solutions 
\eqref{tanh} and \eqref{coth}, and the non-vacuum constant solution 
$\phi(x) = -\varepsilon v$ form a continuously deformable one-parameter
family of degenerate solutions obtained by changing the integration 
constant $c$.

\begin{figure}[th!]
\centering
\subfigure[ ]
{\includegraphics[width=7.8cm]{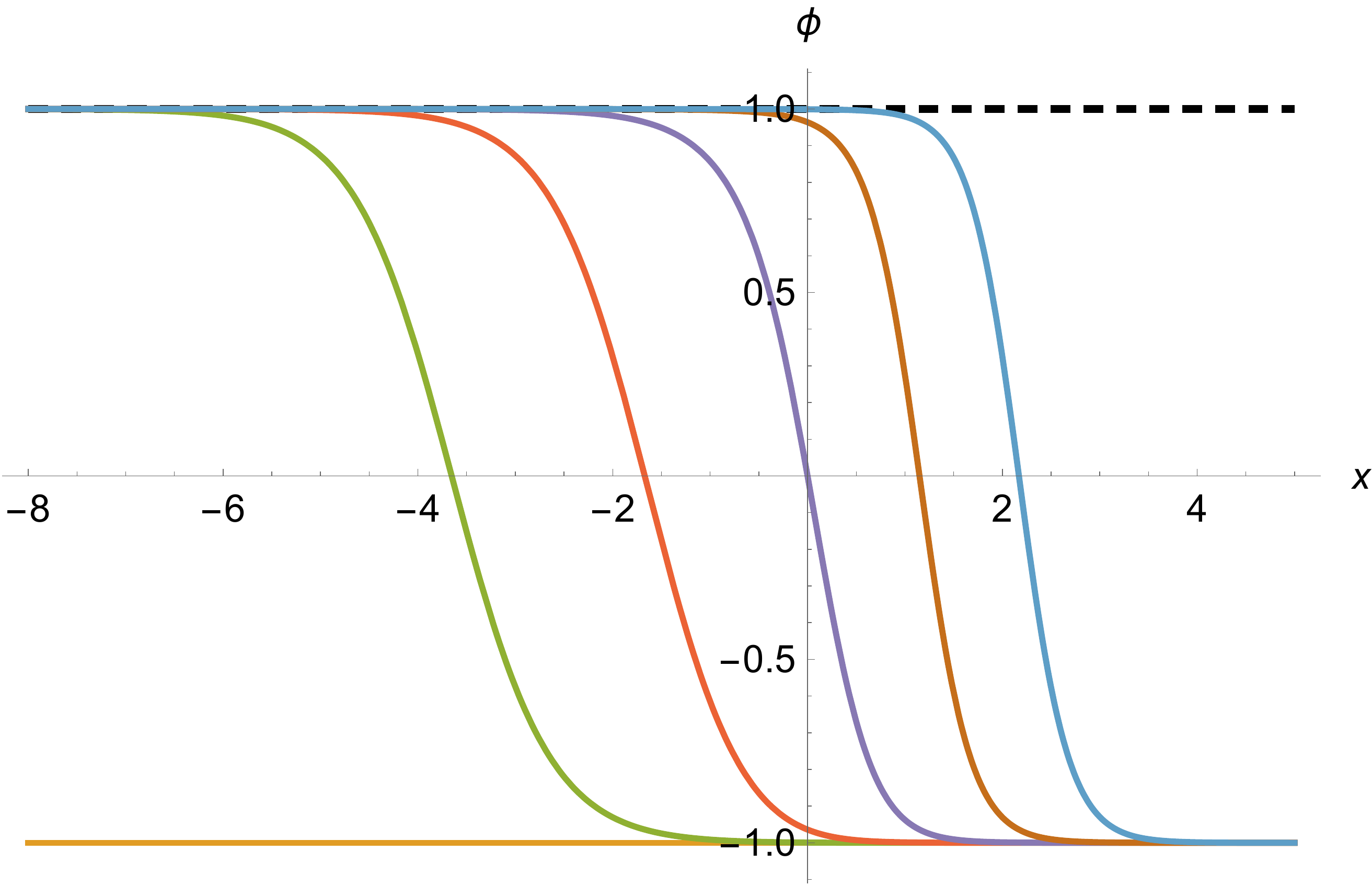}  
\label{fig:phi41}
}
\subfigure[ ]
{\includegraphics[width=7.8cm]{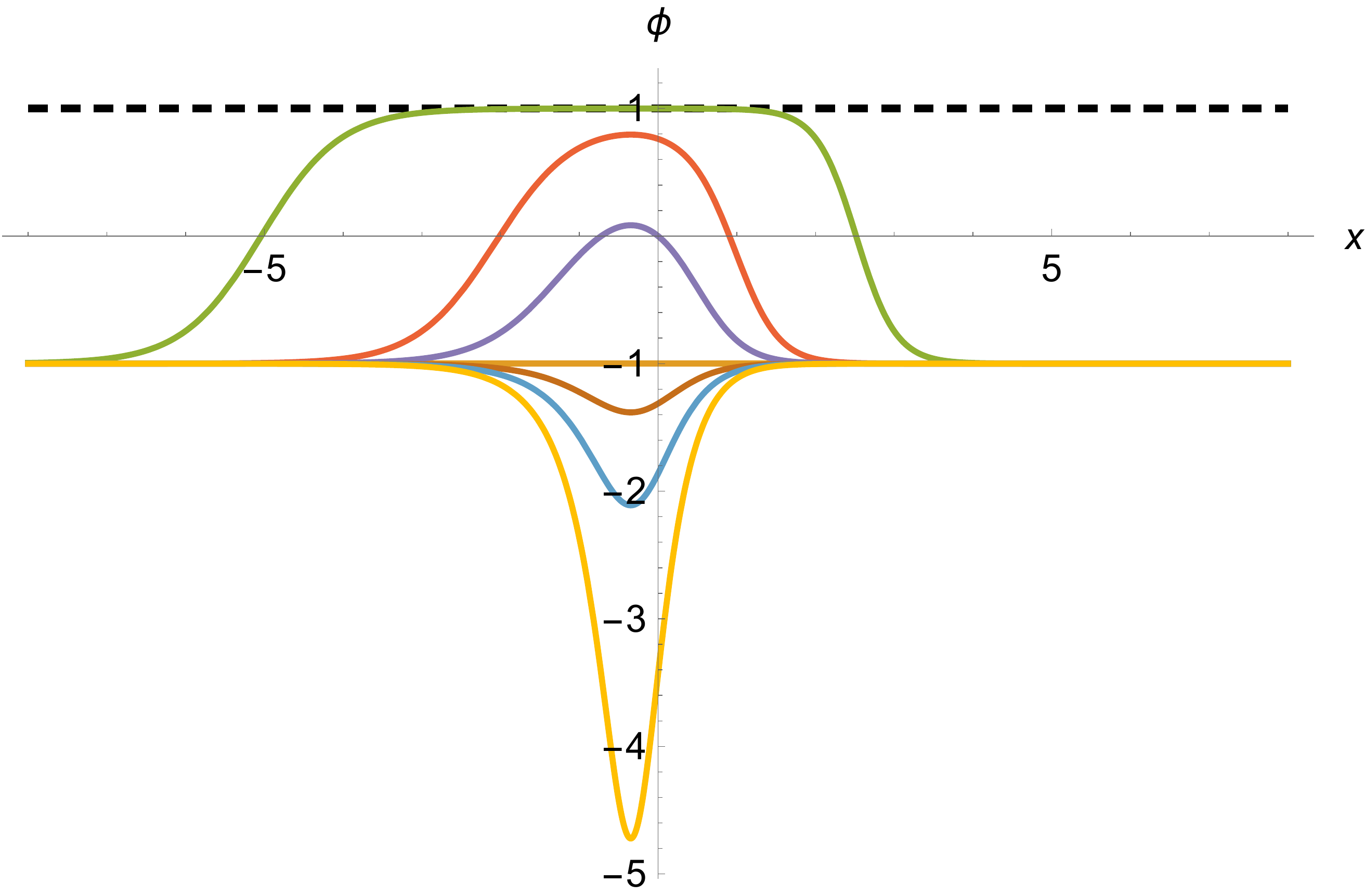}  
\label{fig:phi42}
}
       
\caption{Supersymmetric solutions of deformed $\phi^4$ theory 
for $g(x)$ in \eqref{phi4gx1} with $v=a=1$. 
(a) $g_L=1$ and $g_R=2$. The vacuum is the dashed straight line 
$\phi=1$ with energy $E=-2/3$. The other constant solution $\phi=-1$ has
energy $E=2/3$. Non-constant solutions are for $c=-4,\ -2,\ 0,\ 2$ and 4,
respectively from the left. They all have degenerate energy $E_\textrm{nc}=2$.
(b) $g_L=-1$ and $g_R=2$. The vacuum is the dashed straight line $\phi=1$ 
with energy $E=-2$. The other constant solution $\phi=-1$ has
energy $E=2$. Non-constant solutions are for $c=4,\ 1,\ 0$ in
\eqref{tanh}, and $c= -1,\ -0.6,\ -0.3$ in \eqref{coth}, respectively from 
the top. They all have degenerate energy $E_\textrm{nc}=2$. }
\label{fig:phi4}
\end{figure}

These solutions are illustrated in Fig.~\ref{fig:phi4} for the inhomogeneous 
scaling function
\begin{equation} \label{phi4gx1}
g(x) = \frac12(g_R + g_L)
        + \frac12(g_R - g_L) \tanh a x, \qquad (a >0).
\end{equation}
Note that, compared with the SSG case, we have only antikinks in
Fig.~\ref{fig:phi41}. Also the coth-type solution can indefinitely 
be negative in Fig.~\ref{fig:phi42}, which has also been discussed
in~\cite{Adam:2019djg}.

\end{document}